\documentclass[10pt,a4paper]{article}
\usepackage{amsmath}
\usepackage{amsfonts}
\usepackage{amssymb}
\usepackage[dvips]{graphicx}
\usepackage{bm}

\begin{document}

\title{\textbf{Mass inflation in $f(R)$ gravity}\\ \large{\textsf{A conjecture on the resolution of the mass inflation singularity}}}
\author{\textsc{Dong-il Hwang}$^{a}$\footnote{enotsae@gmail.com},\; \textsc{Bum-Hoon Lee}$^{b}$\footnote{bhl@sogang.ac.kr},\; 
and \textsc{Dong-han Yeom}$^{b,c}$\footnote{innocent.yeom@gmail.com}\\
\textit{$^{a}$\small{Department of Physics, KAIST, Daejeon 305-701, Republic of Korea}}\\
\textit{$^{b}$\small{Center for Quantum Spacetime, Sogang University, Seoul 121-742, Republic of Korea}}\\
\textit{$^{c}$\small{Research Institute for Basic Science, Sogang University, Seoul 121-742, Republic of Korea}}}
\maketitle

\begin{abstract}
We study gravitational collapse of a charged black hole in $f(R)$ gravity using double-null formalism. We require cosmological stability to $f(R)$ models; we used the Starobinsky model and the $R+(1/2)cR^{2}$ model. Charged black holes in $f(R)$ gravity can have a new type of singularity due to higher curvature corrections, the so-called $f(R)$-induced singularity, although it is highly model-dependent. As the advanced time increases, the internal structure will approach the Cauchy horizon, which may not be an inner apparent horizon.

There is mass inflation as one approaches the Cauchy horizon and hence the Cauchy horizon may be a curvature singularity with nonzero area. However, the Ricci scalar is finite for an out-going null observer. This can be integrated as follows: Cosmologically stable higher curvature corrections of the Ricci scalar made it bounded even in the presence of mass inflation.

Finally, we conjecture that if there is a general action including general higher curvature corrections with cosmological stability, then the corrections can make all curvature components finite even in the presence of mass inflation. This might help us to resolve the problem of inner horizon instability of regular black hole models.
\end{abstract}

\newpage

\tableofcontents

\newpage

\section{Introduction}

Quantization of gravity is the final goal of modern physics. To quantize gravity, there exists two non-trivial difficult problems: \textit{gravity is not renormalizable} and \textit{gravity allows various singularities}. In this paper, we try to combine these two problems in an approximated way.

Traditionally, it is well-known that gravity is difficult to quantize due to the problem of renormalization. When we calculate scattering amplitudes for a given action, we have to draw and calculate Feynman diagrams. If a Feynman diagram has a `loop', then the contribution of the diagram simply diverges. By renormalizing parameters suitably, we can cancel infinities and can fit with the observable scattering amplitudes. The problem of quantization of gravity lies in the very fact that, gravity is non-renormalizable. This implies that if the related energy scale is greater than the Planck scale, and hence if the effect of quantum physics and gravity are of similar orders of magnitude, the effective action then has, not only finite terms, but also infinitely many terms to cancel infinities up to leading order. Therefore, such a theory cannot give observable results when the energy scale becomes greater than the Planck scale.

The problem of renormalization may be resolved if we accept string theory. It is an well-established fact that scattering amplitudes of string theory is naturally finite up to all orders. In the low energy limit, string theory gives effective action in the form of supergravity. If we want to know higher order terms of the low energy effective action, we can perturbatively expand it and evaluate all coupling parameters from string theoretical calculations in principle. For example, the effective action of the gravity sector will have the following form \cite{Gasperini:2007zz}:
\begin{eqnarray}
S &=& \frac{1}{2\lambda_{s}^{d-1}} \int dx^{d+1} \sqrt{-g} e^{-\phi} \left[\left( R + (\nabla \phi)^{2} \right) \right.\nonumber \\
&+& \alpha' \left( a_{1} R_{\mu\nu\rho\sigma}^{2} + a_{2} R_{\mu\nu}^{2} + a_{3} R^{2} \right.\nonumber \\
&&+ \left. a_{4} R^{\mu\nu}\nabla_{\mu}\phi\nabla_{\nu}\phi + a_{5} R(\nabla \phi)^{2} + a_{6} R\nabla^{2}\phi + a_{7} (\nabla^{2}\phi)^{2} + a_{8} \nabla^{2}\phi(\nabla\phi)^{2} + a_{9} (\nabla \phi)^{4} \right)\nonumber \\
&+& \left. \alpha'^{2} \mathcal{O}(R^{3}) \right] \nonumber \\
&+& \left[ R + ... \right] + e^{\phi} \left[R + ...\right] + ... \;\; ,
\end{eqnarray}
where $\lambda_{s}$ is the string length, $\alpha'$ is the string tension parameter, $d$ is the space dimension, $\phi$ is the dilaton, and $a_{i}$s are numerical coefficients. Here, what we want to say is that it is too difficult to deal with this general problem without any approximation. Therefore, we have to reduce this action for meaningful calculations.

Let us assume $d=3$ for simplicity. Any reasonable model of string theory should be reduced to a four-dimensional model. In this paper, we will not consider dilaton and other sectors for simplicity; we will only consider the gravity sector including a $U(1)$ gauge field and a complex scalar field to study a charged black hole. We have to choose how to include the higher curvature terms. There can be infinitely many combinations of higher curvature terms: $R$, $A \equiv R_{\mu\nu\rho\sigma}^{2}$, $B \equiv R_{\mu\nu}^{2}$, $C \equiv R_{\mu\nu\rho\sigma}R^{\mu\nu}R^{\rho\sigma}$, etc. The action should have the following general form:
\begin{eqnarray}\label{eq:general}
S = \frac{1}{16\pi} \int dx^{4} \sqrt{-g} \left[ f(R) + f_{2}(A) + f_{3}(B) + f_{4}(C) + ...\;\;\right],
\end{eqnarray}
where $f_{i}$s are arbitrary functions.

At this point, it is important to stress on some general results of the $f(R)$ gravity. If we choose only the $f(R)$ term, then the above action can be changed to that of the Brans-Dicke \cite{Brans:1961sx} type by a simple field redefinition \cite{Magnano:1993bd}. Here, the Brans-Dicke field is related to a non-trivial potential and this potential has further cosmological implications. To make the model cosmologically viable, we have to impose some conditions on the function $f$ \cite{Sotiriou:2008rp}. Imposing such constraints, it is well-known that $f(R)$ gravity can be a viable model of cosmology up to the status of current observations.

Of course, there exist many difficult theoretical problems in this context. Some of them are as follows: (1) What will be the back reactions of higher curvature terms, which were neglected by $f(R)$ gravity? (2) Is the model really renormalizable? (3) Can the model be embedded into string theory? They are difficult to answer. However, at least we can say the two things for general viable $f(R)$ models as follows:
\begin{itemize}
\item Introducing an auxiliary field, we can redefine the gravity sector to be (non-minimally coupled scalar field) + (Ricci scalar) + (non-trivial potential of the auxiliary field).
\item The non-trivial potential should allow a stable minimum to satisfy cosmological constraints.
\end{itemize}
It is not known whether we can generalize these two things for the general action (Equation~(\ref{eq:general})), although we think that it will not be too exotic generalization.

In this paper, we argue that the two conditions for $f(R)$ gravity is useful to resolve a type of singularity -- \textit{the mass inflation singularity}. From the singularity theorem, we know that, if we assume a reasonable causality condition (e.g., global hyperbolicity), a reasonable energy condition (e.g., the null energy condition), and a reasonable initial condition (e.g., a trapping horizon), then there should be a singularity \cite{Hawking:1973uf}. There are some known ways to resolve such singularities; violating some of the assumptions of the singularity theorem, or assuming a certain quantum gravitational correction. These models are known as regular black holes \cite{Frolov:1988vj}\cite{Hossenfelder:2009fc}. However, one of the difficult problem in case of regular black hole models is, not only do they have the outer apparent horizon, but also, they do have an inner apparent horizon. The introduction of the inner apparent horizon is useful in resolving the problem of central singularity. However, the inner horizon is in general unstable via mass inflation \cite{Poisson:1990eh}\cite{Brown:2011tv}. In other words, some of the curvature functions diverge at the inner horizon, and then the model easily goes beyond the range of validity (e.g., Planck scale) of the initial theory. As it stands, there is no known mechanism to resolve such a mass inflation singularity. Perhaps, in case of large number of massless fields, we can trust this model beyond the inner horizon \cite{Yeom:2009zp}. However, for a realistic application, it is difficult. This might imply that, a regular black hole `metric' description always fails, the central region of a black hole is fuzzy and hence we need full quantum gravitational description \cite{Ashtekar:2005cj}.

However, higher curvature corrections may shed some light to the problem. Higher curvature corrections may be helpful in regularizing the mass inflation singularity. In this paper, we assume cosmologically stable $f(R)$ gravity models and study gravitational collapse of a charged black hole. There \textit{is} mass inflation and some of the curvature components may diverge. However, we observe that \textit{the Ricci scalar becomes finite even near the Cauchy horizon}. This feature is derived from the two properties: we include higher order curvature corrections of the Ricci scalar and we impose the cosmological stability. Perhaps, we may generalize that general higher curvature corrections to a cosmologically stable universe can make \textit{all orders} of curvatures to be finite near the Cauchy horizon of a charged black hole. Then, the initial action can be still valid even in the presence of mass inflation. This conjecture will be helpful to understand the mass inflation singularity.

This paper is organized as follows:
In Section~\ref{sec:model}, we describe a model for $f(R)$ gravity and implement the model to double-null formalism to study gravitational collapse.
In Section~\ref{sec:Grav}, we report gravitational collapse in $f(R)$ gravity and discuss mass inflation, and in Section~\ref{sec:dis}, we interpret and discuss our results.

\section{\label{sec:model}Double-null formalism for $f(R)$ gravity}

\subsection{$f(R)$ gravity}

The action of $f(R)$ gravity can be written as
\begin{eqnarray}
S = \int dx^{4} \sqrt{-g} \left[ \frac{1}{16\pi} f(R) + \mathcal{L}_{\mathrm{matter}} \right].
\end{eqnarray}
Introducing an auxiliary field $\psi$, we can change the gravity sector by
\begin{eqnarray}
S_{\mathrm{gravity}} = \frac{1}{16\pi} \int dx^{4} \sqrt{-g} \left[f(\psi) + f'(\psi) (R-\psi) \right]
\end{eqnarray}
with a constraint $\psi = R$. If we define a new field $\Phi$ by
\begin{eqnarray}
\Phi = f'(\psi),
\end{eqnarray}
then we can rewrite the action as
\begin{eqnarray}
S_{\mathrm{gravity}} = \frac{1}{16\pi} \int dx^{4} \sqrt{-g} \left[\Phi R - V(\Phi) \right],
\end{eqnarray}
where
\begin{eqnarray}
V(\Phi) = - f(\psi) + \psi f'(\psi).
\end{eqnarray}
This is exactly the $\omega = 0$ limit of the Brans-Dicke theory.

\subsection{\label{sec:The}Brans-Dicke theory}

The action of the Brans-Dicke theory with a scalar field becomes
\begin{eqnarray}\label{eq:BDscalar}
S_{\mathrm{BD}} &=& \int dx^{4} \sqrt{-g} \left[ \frac{1}{16\pi} \left( \Phi R - \frac{\omega}{\Phi}\Phi_{;\mu}\Phi_{;\nu}g^{\mu\nu} - V(\Phi) \right) \right. \nonumber \\
&& + \left. \left(- \frac{1}{2}\left(\phi_{;\mu}+ieA_{\mu}\phi \right)g^{\mu\nu}\left(\overline{\phi}_{;\nu}-ieA_{\nu}\overline{\phi}\right)-\frac{1}{16\pi}F_{\mu\nu}F^{\mu\nu} \right) \right],
\end{eqnarray}
where $R$ is the Ricci scalar, $\Phi$ is the Brans-Dicke field, and $\phi$ is a complex scalar field with a gauge coupling $e$ and a gauge field $A_{\mu}$, where $F_{\mu\nu}=A_{\nu;\mu}-A_{\mu;\nu}$.
Here, $\omega$ is the Brans-Dicke coupling constant and we choose $\omega=0$.

The Einstein equation is as follows:
\begin{eqnarray}\label{eq:Einstein}
G_{\mu\nu} = 8 \pi T^{\mathrm{BD}}_{\mu\nu} + 8 \pi \frac{T^{\mathrm{C}}_{\mu\nu}}{\Phi} \equiv 8 \pi T_{\mu\nu},
\end{eqnarray}
where the Brans-Dicke part of the energy-momentum tensor is
\begin{eqnarray}\label{eq:T_BD}
T^{\mathrm{BD}}_{\mu\nu} = \frac{1}{8\pi \Phi} \left(-g_{\mu\nu}\Phi_{;\rho \sigma}g^{\rho\sigma}+\Phi_{;\mu\nu}\right)
+ \frac{\omega}{8\pi \Phi^{2}} \left(\Phi_{;\mu}\Phi_{;\nu}-\frac{1}{2}g_{\mu\nu}\Phi_{;\rho}\Phi_{;\sigma}g^{\rho\sigma}\right) - g_{\mu\nu}\frac{V(\Phi)}{16 \pi}
\end{eqnarray}
and the matter part of the energy-momentum tensor is
\begin{eqnarray}\label{eq:T_C}
T^{\mathrm{C}}_{\mu\nu} &=& \frac{1}{2}\left(\phi_{;\mu}\overline{\phi}_{;\nu}+\overline{\phi}_{;\mu}\phi_{;\nu}\right)
\nonumber \\
&& {}+\frac{1}{2}\left(-\phi_{;\mu}ieA_{\nu}\overline{\phi}+\overline{\phi}_{;\nu}ieA_{\mu}\phi+\overline{\phi}_{;\mu}ieA_{\nu}\phi-\phi_{;\nu}ieA_{\mu}\overline{\phi}\right)
\nonumber \\
&& {}+\frac{1}{4\pi}F_{\mu \rho}{F_{\nu}}^{\rho}+e^{2}A_{\mu}A_{\nu}\phi\overline{\phi}+\mathcal{L}^{\mathrm{EM}}g_{\mu \nu}.
\end{eqnarray}

The field equations are as follows:
\begin{eqnarray}
\label{eq:Phi1}\Phi_{;\mu\nu}g^{\mu\nu}-\frac{8\pi}{3+2\omega}T^{\mathrm{C}}-\frac{1}{3+2\omega}\left( \Phi \frac{dV}{d\Phi} - 2 V \right) &=&0, \\
\label{eq:phi}\phi_{;\mu\nu}g^{\mu\nu}+ieA^{\mu}\left(2\phi_{;\mu}+ieA_{\mu}\phi\right)+ieA_{\mu;\nu}g^{\mu\nu}\phi &=& 0,
\\
\label{eq:a}\frac{1}{2\pi} {F^{\nu}}_{\mu;\nu} -ie\phi\left(\overline{\phi}_{;\mu}-ieA_{\mu}\overline{\phi}\right)+ie\overline{\phi}\left(\phi_{;\mu}+ieA_{\mu}\phi\right) &=& 0,
\end{eqnarray}
where
\begin{eqnarray}\label{eq:T}
T^{\mathrm{C}} = {T^{\mathrm{C}}}^{\mu}_{\;\mu}.
\end{eqnarray}

\subsection{\label{sec:Cho}Choice of potential $V(\Phi)$}

In order to make a $f(R)$ model free from ghost, we consider $\Phi = f'(R) > 0$. Additionally, for the consistency of the field redefinition, we only consider a region $f''(R) > 0$. Dynamics of the Brans-Dicke field is governed by $\Phi V'(\Phi) - 2 V(\Phi) = 2f(\psi) - \psi f'(\psi)$. In other words, if $T^{\mathrm{C}} = 0$ and $\Phi V'(\Phi) - 2 V(\Phi) = 2f(\psi) - \psi f'(\psi) = 0$, there can be an extreme point or a constant curvature point $R = R_{0}$. For convenience, we define an effective potential $U(\Phi)$ as
\begin{eqnarray}
U(\Phi) = \int^{\Phi} \left( \bar{\Phi} V'(\bar{\Phi}) - 2V(\bar{\Phi}) \right) d \bar{\Phi}.
\end{eqnarray}
Then the equation of the Brans-Dicke field becomes:
\begin{eqnarray}
\Phi_{;\mu\nu}g^{\mu\nu}-\frac{8\pi}{3+2\omega}T^{\mathrm{C}}-\frac{1}{3+2\omega} \frac{dU}{d\Phi} =0
\end{eqnarray}
and the stability around the extreme point becomes
\begin{eqnarray}
\left. \frac{d^{2} U}{d \Phi^{2}}\right|_{\Phi_{0}} = \left. \left( \frac{f'(\psi)}{f''(\psi)} - \psi \right) \right|_{\psi_{0}} > 0,
\end{eqnarray}
where $\Phi_{0}$ and $\psi_{0}$ are solutions of $\Phi V'(\Phi) - 2 V(\Phi) = 2f(\psi) - \psi f'(\psi) = 0$.

\subsection{\label{sec:imp}Implementation of the double-null formalism}

We use the double-null coordinates
\begin{eqnarray}\label{eq:doublenull}
ds^{2} = -\alpha^{2}(u,v) du dv + r^{2}(u,v) d\Omega^{2},
\end{eqnarray}
assuming spherical symmetry. Here, $u$ is the retarded time, $v$ is the advanced time, and $\theta$ and $\varphi$ are the angular coordinates.

We follow the notation of previous papers \cite{Yeom1}\cite{Yeom2}\cite{Yeom3}\cite{Hwang:2010aj}\cite{Avelino:2009vv}: the metric function $\alpha$, the radial function $r$, the Brans-Dicke field $\Phi$, and a scalar field $s \equiv \sqrt{4\pi} \phi$, and define
\begin{eqnarray}\label{eq:conventions}
h \equiv \frac{\alpha_{,u}}{\alpha},\quad d \equiv \frac{\alpha_{,v}}{\alpha},\quad f \equiv r_{,u},\quad g \equiv r_{,v},\quad W \equiv \Phi_{,u},\quad Z \equiv \Phi_{,v}, \quad w \equiv s_{,u},\quad z \equiv s_{,v}.
\end{eqnarray}

The Einstein tensors are then expressed as follows:
\begin{eqnarray}
\label{eq:Guu}G_{uu} &=& -\frac{2}{r} \left(f_{,u}-2fh \right),\\
\label{eq:Guv}G_{uv} &=& \frac{1}{2r^{2}} \left( 4 rf_{,v} + \alpha^{2} + 4fg \right),\\
\label{eq:Gvv}G_{vv} &=& -\frac{2}{r} \left(g_{,v}-2gd \right),\\
\label{eq:Gthth}G_{\theta\theta} &=& -4\frac{r^{2}}{\alpha^{2}} \left(d_{,u}+\frac{f_{,v}}{r}\right).
\end{eqnarray}

Also, we can obtain the energy-momentum tensor for the Brans-Dicke field part and the scalar field part ($\omega=0$):
\begin{eqnarray}
\label{eq:TBDuu}T^{\mathrm{BD}}_{uu} &=& \frac{1}{8 \pi \Phi} (W_{,u} - 2hW), \\
\label{eq:TBDuv}T^{\mathrm{BD}}_{uv} &=& - \frac{Z_{,u}}{8 \pi \Phi} - \frac{gW+fZ}{4 \pi r \Phi} + \frac{\alpha^{2}V}{32 \pi \Phi},\\
\label{eq:TBDvv}T^{\mathrm{BD}}_{vv} &=& \frac{1}{8 \pi \Phi} (Z_{,v} - 2dZ), \\
\label{eq:TBDthth}T^{\mathrm{BD}}_{\theta\theta} &=& \frac{r^{2}}{2 \pi \alpha^{2} \Phi} Z_{,u} + \frac{r}{4 \pi \alpha^{2} \Phi} (gW+fZ) - \frac{r^{2}V}{16 \pi \Phi},
\end{eqnarray}
\begin{eqnarray}
\label{eq:TSuu}T^{\mathrm{C}}_{uu} &=& \frac{1}{4\pi} \left[ w\overline{w} + iea(\overline{w}s-w\overline{s}) +e^{2}a^{2}s\overline{s} \right],\\
\label{eq:TSuv}T^{\mathrm{C}}_{uv} &=& \frac{{(a_{,v})}^{2}}{4\pi\alpha^{2}},\\
\label{eq:TSvv}T^{\mathrm{C}}_{vv} &=& \frac{1}{4\pi} z\overline{z},\\
\label{eq:TSthth}T^{\mathrm{C}}_{\theta\theta} &=& \frac{r^{2}}{4\pi\alpha^{2}} \left[ (w\overline{z}+z\overline{w}) + iea(\overline{z}s-z\overline{s})+\frac{2{(a_{,v})}^{2}}{\alpha^{2}} \right],
\end{eqnarray}

To implement the double-null formalism into the numerical scheme, it is convenient to represent all the equations as first order differential equations.
Note that
\begin{eqnarray}\label{eq:T_trace}
T^{\mathrm{C}} = - \frac{4}{\alpha^{2}}T^{\mathrm{C}}_{uv} + \frac{2}{r^{2}}T^{\mathrm{C}}_{\theta \theta}
\end{eqnarray}
and
\begin{eqnarray}
\Phi_{;\mu\nu}g^{\mu\nu} = - \frac{4}{r\alpha^{2}} \left( r\Phi_{,uv} + r_{,v}\Phi_{,u} + r_{,u}\Phi_{,v} \right).
\end{eqnarray}

The Einstein equations for $\alpha_{,uv}$, $r_{,uv}$, and the field equation for $\Phi$ are then coupled:
\begin{eqnarray}\label{eq:coupled}
\left( \begin{array}{ccc}
1 & 1/r & 1/\Phi \\
0 & 1 & r/2\Phi \\
0 & 0 & r
\end{array} \right)
\left( \begin{array}{c}
d_{,u} \\
f_{,v} \\
Z_{,u}
\end{array} \right)
= \left( \begin{array}{c}
\mathfrak{A} \\
\mathfrak{B} \\
\mathfrak{C}
\end{array} \right)
\end{eqnarray}
where
\begin{eqnarray}
\label{eq:A}\mathfrak{A} &\equiv&-\frac{2\pi \alpha^{2}}{r^{2}\Phi}T^{\mathrm{C}}_{\theta\theta} - \frac{1}{2r}\frac{1}{\Phi}(gW+fZ) + \frac{\alpha^{2}V}{8 \Phi}, \\
\label{eq:B}\mathfrak{B} &\equiv& \frac{4 \pi r}{\Phi}T^{\mathrm{C}}_{uv} - \frac{\alpha^{2}}{4r} - \frac{fg}{r} - \frac{1}{\Phi}(gW+fZ) + \frac{r \alpha^{2}}{8 \Phi}V, \\
\label{eq:C}\mathfrak{C} &\equiv& - fZ - gW - \frac{4\pi r}{3} \left(\frac{\alpha^{2}}{2}T^{\mathrm{C}} + \frac{\alpha^{2}}{16 \pi}\left(\Phi V' - 2 V \right) \right).
\end{eqnarray}

After solving these coupled system of equations, we can write all the equations as follows:
\begin{eqnarray}
\label{eq:E1}f_{,u} &=& 2fh - \frac{r}{2 \Phi} (W_{,u}-2hW) - \frac{4 \pi r}{\Phi} T^{\mathrm{C}}_{uu} ,\\
\label{eq:E2}g_{,v} &=& 2gd - \frac{r}{2 \Phi} (Z_{,v}-2dZ) - \frac{4 \pi r}{\Phi} T^{\mathrm{C}}_{vv} ,\\
\label{eq:E3}d_{,u} = h_{,v} &=& \mathfrak{A} - \frac{\mathfrak{B}}{r} - \frac{\mathfrak{C}}{2r\Phi},\\
\label{eq:E4}g_{,u} = f_{,v} &=& \mathfrak{B} - \frac{\mathfrak{C}}{2\Phi},\\
\label{eq:Phi}Z_{,u} = W_{,v} &=& \frac{\mathfrak{C}}{r},
\end{eqnarray}
including the scalar field equation
\begin{eqnarray}
\label{eq:fieldeqns1}a_{,v} &=& \frac{\alpha ^{2} q}{2 r^{2}}, \\
\label{eq:fieldeqns2}q_{,v} &=& -\frac{ier^{2}}{2} (\overline{s}z-s\overline{z}), \\
\label{eq:fieldeqns3}z_{,u} = w_{,v} &=& - \frac{fz}{r} - \frac{gw}{r} - \frac{iearz}{r} - \frac{ieags}{r} - \frac{ie}{4r^{2}}\alpha^{2}qs.
\end{eqnarray}

In fact, for a given $f(R)$, we have to find the inverse function of $f'(R)$ to write $V$ as a function of $\Phi$. In general, it is not convenient for a numerical calculation. Therefore, we will change terms depending on $V(\Phi)$ with the help of the following identities:
\begin{eqnarray}
V(\Phi) &=& -f(R) + R f'(R),\\
\Phi \frac{dV}{d\Phi} - 2 V &=& 2f(R) - R f'(R),
\end{eqnarray}
where the Ricci scalar in the double-null coordinates can be expressed as
\begin{eqnarray}\label{eq:Ricci}
R = \frac{2}{\alpha^{2}} \left[ 4 \left(\frac{\alpha_{,u}}{\alpha}\right)_{,v} + 8 \frac{r_{,uv}}{r} + \frac{\alpha^{2}}{r^{2}} + \frac{4r_{,u}r_{,v}}{r^{2}} \right].
\end{eqnarray}
In addition to this, we use the following two equations to control the evolution of Ricci scalar $R$:
\begin{eqnarray}
\label{eq:R1}R_{,u} &=& \frac{\Phi_{,u}}{f''(R)},\\
\label{eq:R2}R_{,v} &=& \frac{\Phi_{,v}}{f''(R)}.
\end{eqnarray}

Now, the equations for $\alpha_{,uv}$, $r_{,uv}$, $\Phi_{,uv}$, and $s_{,uv}$ parts can be represented as first order differential equations.
We can then implement the same integration scheme as was used in previous papers \cite{Yeom1}\cite{Yeom2}\cite{Yeom3}\cite{Hwang:2010aj} to solve the Brans-Dicke theory.
We use the second order Runge-Kutta method to solve the problem \cite{nr}. Tests of the convergence are provided in the Appendix.

\subsection{\label{sec:ini}Initial conditions and free parameters}

We need initial conditions for all the functions ($\alpha, h, d, r, f, g, \Phi, W, Z, s, w, z, a, q, R$) at the initial $u=u_{\mathrm{i}}$ and $v=v_{\mathrm{i}}$ surfaces, where we set $u_{\mathrm{i}}=v_{\mathrm{i}}=0$.

We have the gauge freedom to choose the initial $r$ function. Although all the constant $u$ and $v$ lines are null, the freedom to choose the distances between them still remains. Here, we choose $r(0,0)=r_{0}$, $f(u,0)=r_{u0}$, and $g(0,v)=r_{v0}$, where $r_{u0}<0$ and $r_{v0}>0$ such that the radial function for an in-going observer decreases and that for an out-going observer increases.

First, we assume that the Brans-Dicke field is asymptotically $\Phi_{0}$, where $\Phi_{0} V'(\Phi_{0}) - 2 V(\Phi_{0}) = 2 f(R_{0})-R_{0}f'(R_{0})=0$. Then, $\Phi(u,0)=\Phi(0,v)=\Phi_{0}$, $W(u,0)=Z(0,v)=0$, and $R(u,0)=R(0,v)=R_{0}$. We use a shell-shaped scalar field, and hence its interior is not affected by the shell.
Thus, we can simply choose $s(u,0)=0$. Also, $w(u,0)=h(u,0)=a(u,0)=q(a,0)=0$ holds.
Since the asymptotic mass function
\begin{eqnarray}
m(u,v) \equiv \frac{r}{2}\left( 1+4\frac{r_{,u}r_{,v}}{\alpha^{2}} + \frac{q^{2}}{r^{2}} - \frac{V}{6\Phi}r^{2} \right)
\end{eqnarray}
should vanish at $u=v=0$, it is convenient to choose $r_{u0}=-1/2$ and $r_{v0}=1/2$, and then
\begin{eqnarray}
\alpha(0,0)=\left(1-\frac{V(\Phi_{0})}{6\Phi_{0}} r_{0}^{2}\right)^{-1/2}.
\end{eqnarray}

We need more information to determine $d, g, Z$, and $z$ at the $v=0$ surface. We obtain $d$ from Equation~(\ref{eq:E3}), $g$ from Equation~(\ref{eq:E4}), $Z$ from Equation~(\ref{eq:Phi}), and $z$ from Equation~(\ref{eq:fieldeqns3}).

We can choose an arbitrary function for $s(0,v)$ to induce a collapsing pulse. In this paper, we use
\begin{eqnarray} \label{s_initial}
s(u_{\mathrm{i}},v)= A \sin^{2} \left( \pi \frac{v-v_{\mathrm{i}}}{v_{\mathrm{f}}-v_{\mathrm{i}}} \right) \left[ \cos \left( 2 \pi \frac{v-v_{\mathrm{i}}}{v_{\mathrm{f}}-v_{\mathrm{i}}} \right) + i \cos\left( 2 \pi \frac{v-v_{\mathrm{i}}}{v_{\mathrm{f}}-v_{\mathrm{i}}} + \delta \right) \right]
\end{eqnarray}
for $0\leq v \leq v_{\mathrm{f}}$ and $s(0,v)=0$ otherwise, where $v_{\mathrm{f}}$ is the width of the pulse, $A$ is the amplitude, and $\delta$ is a free parameter to tune the phase of the complex scalar field \cite{Yeom2}\cite{Yeom3}.
Then we obtain $z(u_{\mathrm{i}},v)$.
This implements one pulse of energy ($T_{vv} \sim z^{2}$) along the out-going null direction by a differentiable function $z(0,v)$.
Also, from Equation~(\ref{eq:E2}), we can use $d = r|z|^{2}/2g\Phi$ on the $u=0$ surface and thus we obtain $d(0,v)$. By integrating $d$ along $v$, we have $\alpha(0,v)$.

We need more information for $h, f, W, w, a$ and $q$ at the $u=0$ surface. We obtain $h$ from Equation~(\ref{eq:E3}), $f$ from Equation~(\ref{eq:E4}), $W$ from Equation~(\ref{eq:Phi}), $w$ from Equation~(\ref{eq:fieldeqns3}), $a$ from Equation~(\ref{eq:fieldeqns1}), and $q$ from Equation~(\ref{eq:fieldeqns2}). This finishes the task of assignment of initial conditions.

We choose $r_{0}=10$, $e=0.3$, and $\omega=0$, leaving the two parameters $(A, \delta)$ free, where $A$ is the amplitude of the pulse of the scalar field and $\delta$ is the phase of the complex scalar field.

\begin{figure}
\begin{center}
\includegraphics[scale=0.35]{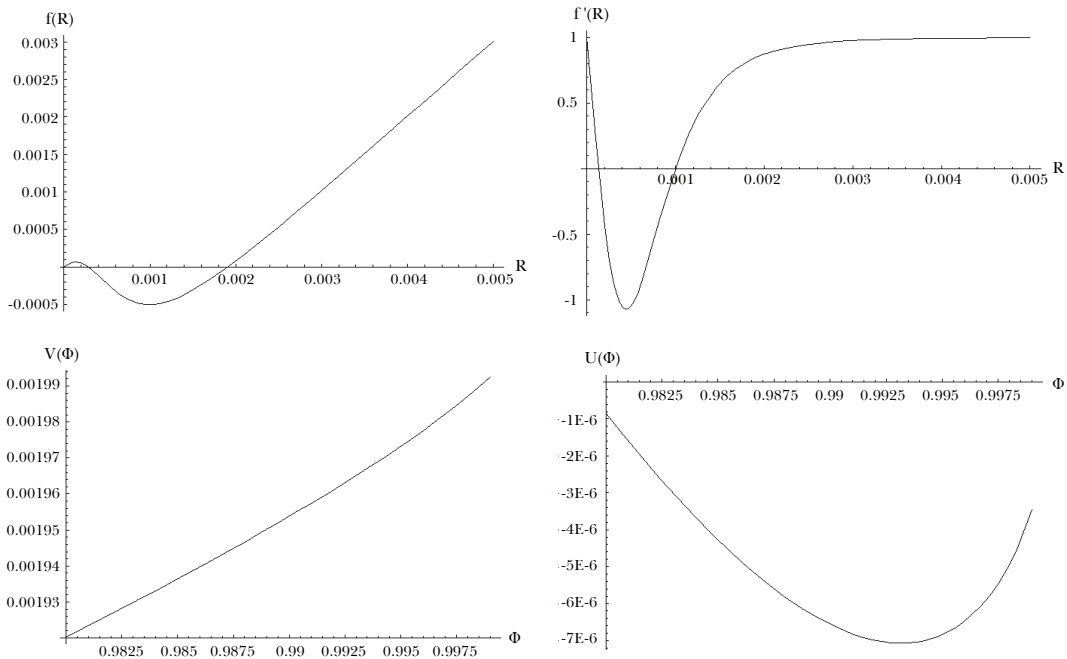}
\caption{\label{fig:Starobinsky}The Starobinsky model with $\lambda=2$, $R_{0}=0.001$, and $n=2$: $f(R)$ (upper left), $f'(R)$ (upper right), $V(\Phi)$ (lower left), and $U(\Phi)$ (lower right).}
\end{center}
\end{figure}

\section{\label{sec:Grav}Gravitational collapse in $f(R)$ gravity}

\subsection{Models and the $f(R)$-induced singularity}

In this paper, we study the following two models.

\subsubsection{Model~$1$: Starobinsky model}

We first use the so-called Starobinsky model \cite{Starobinsky:2007hu}:
\begin{eqnarray}
f(R) = R + \lambda R_{0} \left( \left(1+\frac{R^{2}}{R_{0}^{2}} \right)^{-n}-1 \right),
\end{eqnarray}
where $\lambda$, $R_{0}$, and $n$ are free parameters. In this paper, we choose $\lambda=2$, $R_{0}=0.001$, and $n=2$ (Figure~\ref{fig:Starobinsky}). We can see that there is a stable equilibrium around $\Phi_{0}\simeq0.9932$ and $V_{0}\simeq0.001966$. Therefore, we can choose these initial conditions for $\Phi_{0}$ and $V_{0}$. Then the other initial parameters are determined by consistency condition. In this setup, the cosmological horizon is $l = \sqrt{6\Phi_{0}/V_{0}} \simeq 55$. Therefore, for convenience, we should choose the size of black holes less than $l \simeq 55$.

It should be noted that if $R$ is less than $R_{0} = 0.001$, then the Brans-Dicke field $\Phi=f'(R)$ should be less than zero (upper left of Figure~\ref{fig:Starobinsky}). Therefore, the region of validity for $R$ is $R_{0} < R < \infty$. On the other hand, in this region, the Brans-Dicke field $\Phi$ varies from zero to one. In other words, $\Phi$ cannot be greater than one. This particular behavior is due to the choice of $f(R)$: $f(R) \sim R$ in the large $R$ limit.

Therefore, in the large $R$ limit, we observe that
\begin{eqnarray}
f(R) &\simeq& R - \lambda R_{0} + \lambda R_{0}^{2n+1} \frac{1}{R^{2n}} + ...,\\
f'(R) &\simeq& 1 - 2n \lambda R_{0}^{2n+1} \frac{1}{R^{2n+1}} + ...,\\
f''(R) &\simeq& 2n(2n+1) \lambda R_{0}^{2n+1} \frac{1}{R^{2n+2}} + ...,\\
\lim_{\Phi \rightarrow 1} U(\Phi) &=& \lim_{\Phi \rightarrow 1} \int^{\Phi} \left(\bar{\Phi} \frac{V(\bar{\Phi})}{d \bar{\Phi}} - 2V(\bar{\Phi}) \right) d\bar{\Phi}\\
&=& \lim_{R \rightarrow \infty} \int^{R} \left(2f(\bar{R}) - \bar{R} f'(\bar{R}) \right) f''(\bar{R}) d\bar{R}\\
&\propto& \int^{\infty} \frac{d\bar{R}}{\bar{R}^{2n+1}}.
\end{eqnarray}
Therefore, if $n>0$, then the effective potential $U(\Phi)$ in the $\Phi=1$ limit is finite, although the effective force term $2f - Rf' = \Phi V' -2 V$ is infinite.

Now let us observe the behavior of potentials $V(\Phi)$ and $U(\Phi)$ (lower left and lower right of Figure~\ref{fig:Starobinsky}). Here, we plot $\Phi < 1$. If the Brans-Dicke field $\Phi$ becomes much smaller than the local minimum in terms of the effective potential $U(\Phi)$ (for example, if the field rolls up to $\Phi \sim 0.5$), then it will roll down and oscillate around the local minimum. 
However, it is potentially possible for $\Phi$ to touch the point $\Phi=1$; then the region should be identified with the curvature singularity $R = \infty$. This is a kind of singularity, but not from the point of view of general relativity. Such kind of singularity highly depends on the choice of $f(R)$. We call this kind of singularity the \textit{$f(R)$-induced singularity}. Such a singularity will happen for various $f(R)$ models, depending on the particular case; perhaps, a suitable choice of $f(R)$ may remove such singularities.

\subsubsection{Model~$2$: $R+(1/2)cR^{2}$}

In order to avoid the $f(R)$-induced singularity and to study the general feature of the $f(R)$ gravity, we choose a simpler model that has no $f(R)$-induced singularity. Here, we study the simplest $f(R)$ model:
\begin{eqnarray}
f(R) = R + \frac{1}{2}c R^{2}.
\end{eqnarray}
Here, $c$ is an arbitrary constant. Then
\begin{eqnarray}
f'(\psi) = 1 + c \psi = \Phi
\end{eqnarray}
and
\begin{eqnarray}
\psi = \frac{\Phi - 1}{c}.
\end{eqnarray}
Therefore,
\begin{eqnarray}
V(\Phi) &=& -f(\psi) + \frac{df(\psi)}{d\psi} \psi\\
&=& \frac{1}{2c} \left( \Phi - 1 \right)^{2}
\end{eqnarray}
and
\begin{eqnarray}
U(\Phi) &=& \int \left( \Phi \frac{dV}{d\Phi} - 2V \right) d\Phi \\
&=& \frac{1}{2c} \left( \Phi - 1 \right)^{2}.
\end{eqnarray}
For stability, we choose $c>0$.

\begin{figure}
\begin{center}
\includegraphics[scale=0.3]{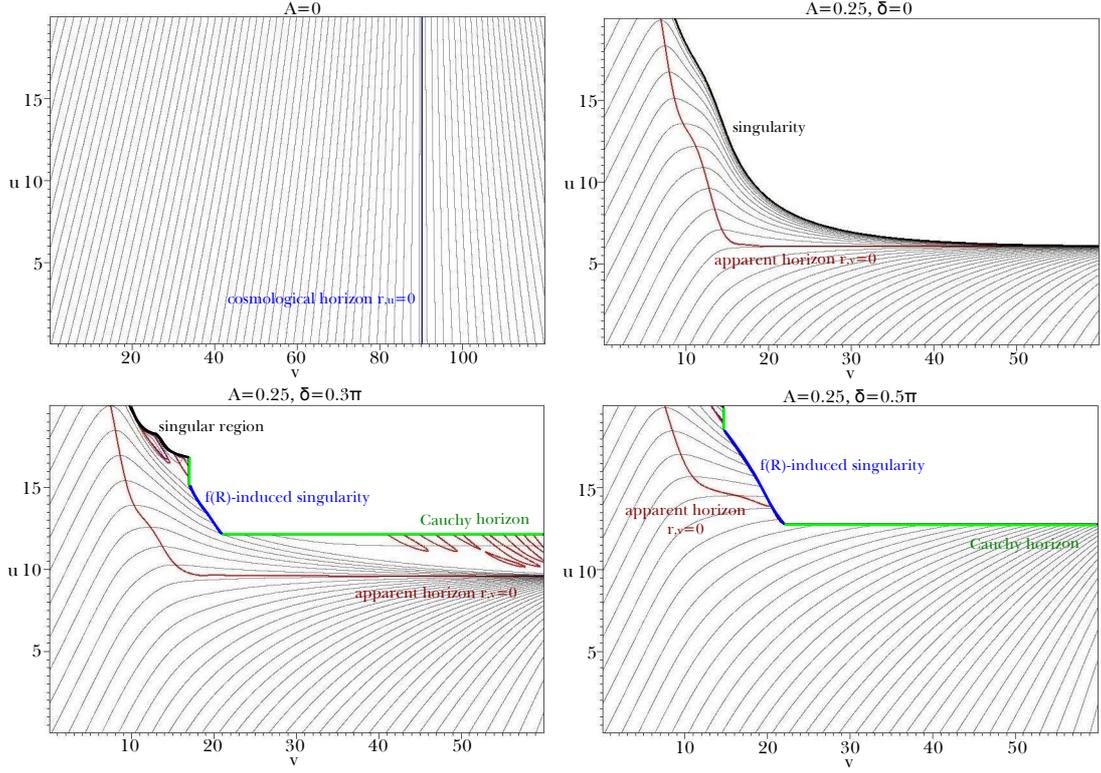}
\caption{\label{fig:Model1}Results of Model~$1$: de Sitter limit (upper left), neutral limit (upper right), extreme limit (lower right), and intermediate limit (lower left).}
\end{center}
\end{figure}
\begin{figure}
\begin{center}
\includegraphics[scale=0.3]{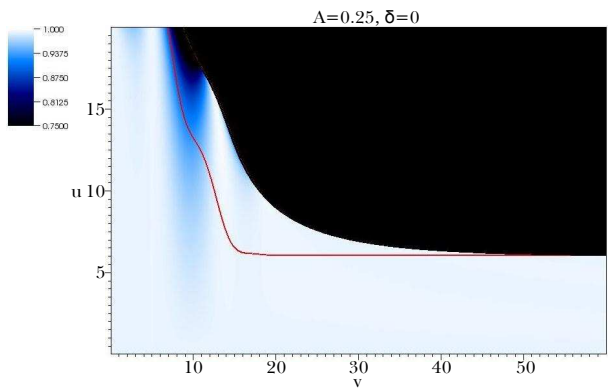}
\caption{\label{fig:neutral_Sb}$\Phi$ of the $A=0.25$ and $\delta=0$ case (Model~$1$).}
\end{center}
\end{figure}

\subsection{Model~$1$: Basic features}

\subsubsection{The simplest case: de Sitter space}

As a triviality test, we investigate the $A=0$ case. Then we observe the de Sitter space without a black hole (upper left of Figure~\ref{fig:Model1}). The cosmological horizon is at $r = l \sim 55$ and parallel to the in-going null direction. Despite being a basic result, this shows a non-trivial consistency check for our simulations.

\subsubsection{Neutral limit}

We now observe a neutral black hole and a less charged black hole (upper right of Figure~\ref{fig:Model1}). First, we observe $A=0.25$ and $\delta=0$. In this case, as there is no phase difference between the real and the imaginary part of the complex scalar field $\phi$, the matter combination becomes neutral. Therefore, we can see just a space-like singularity and an apparent horizon.

Figure~\ref{fig:neutral_Sb} shows the response of the Brans-Dicke field during gravitational collapse. Basically, the Brans-Dicke field first moves towards the lower value instead of the asymptotic value, and then returns to a certain limit. The similar situations have already been studied by the authors in \cite{Hwang:2010aj}. In that paper, it has been observed that the collapse of a neutral matter field induces a decrease of the Brans-Dicke field for $\omega > -3/2$; and as $\omega$ approaches the $-3/2$ limit, the Brans-Dicke field becomes more and more sensitive. The similar behavior can be checked for $f(R)$ gravity as well, since $f(R)$ gravity is the $\omega=0$ limit of the Brans-Dicke theory. Of course, there is one major difference between the present and the previous work. In the previous paper, there is no potential and hence the field value returns to the asymptotic value, while in the present paper, the field value is governed by a potential and hence whole processes is more complex.

\subsubsection{\label{sec:nearextreme}Near extreme limit: $f(R)$-induced singularity}

Here, we observe $A=0.25$ and $\delta=0.5\pi$ (lower right of Figure~\ref{fig:Model1}). In this limit, we can push as much charge as possible \cite{Yeom3}. In the Einstein gravity, because of electromagnetic repulsion, it is impossible to observe a naked singularity or an exact extreme black hole \cite{Yeom1}\cite{Yeom2}\cite{Yeom3}.

\begin{figure}
\begin{center}
\includegraphics[scale=1]{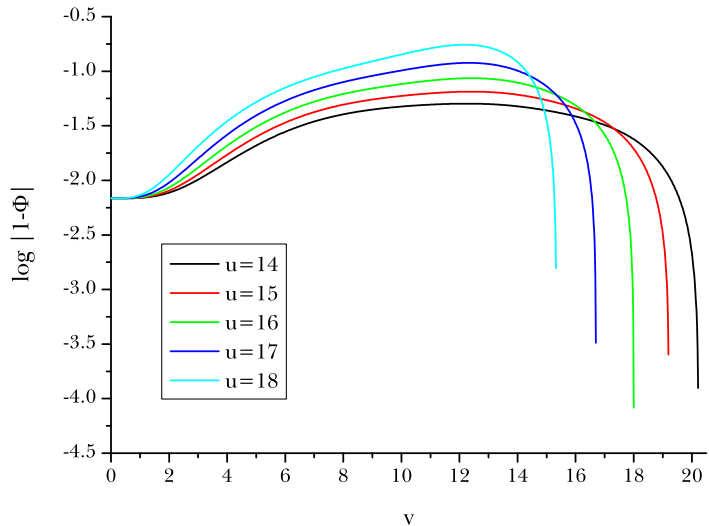}
\caption{\label{fig:f(R)induced}$\log |1-\Phi|$ as a function of $v$ along $u=14, 15, 16, 17, 18$ for the $A=0.25$ and $\delta=0.5 \pi$ case (Model~$1$).}
\end{center}
\end{figure}

However, one interesting point is that the $f(R)$-induced singularity is located outside the apparent horizon. Gravitational collapse will perturb the Brans-Dicke field value to become smaller than the local minimum of the effective potential $U(\Phi)$. Then, the perturbed Brans-Dicke field will roll down and rolls up to approach $\Phi \rightarrow 1$. Figure~\ref{fig:f(R)induced} shows such a behavior around the $f(R)$-induced singularity. Figure~\ref{fig:f(R)induced} shows the behavior of $\log|1-\Phi|$ as a function of $v$. As $v$ increases, $|1-\Phi|$ first increases (i.e., $\Phi$ decreases), and then $|1-\Phi|$ decreases (i.e., $\Phi$ approaches one) and eventually the simulation breaks down.
This explains the behavior of the $f(R)$-induced singularity, since the $\Phi \rightarrow 1$ limit is the $R \rightarrow \infty$ limit for the model. This singularity eventually induces the Cauchy horizons along the in-going and the out-going directions.

\begin{figure}
\begin{center}
\includegraphics[scale=1]{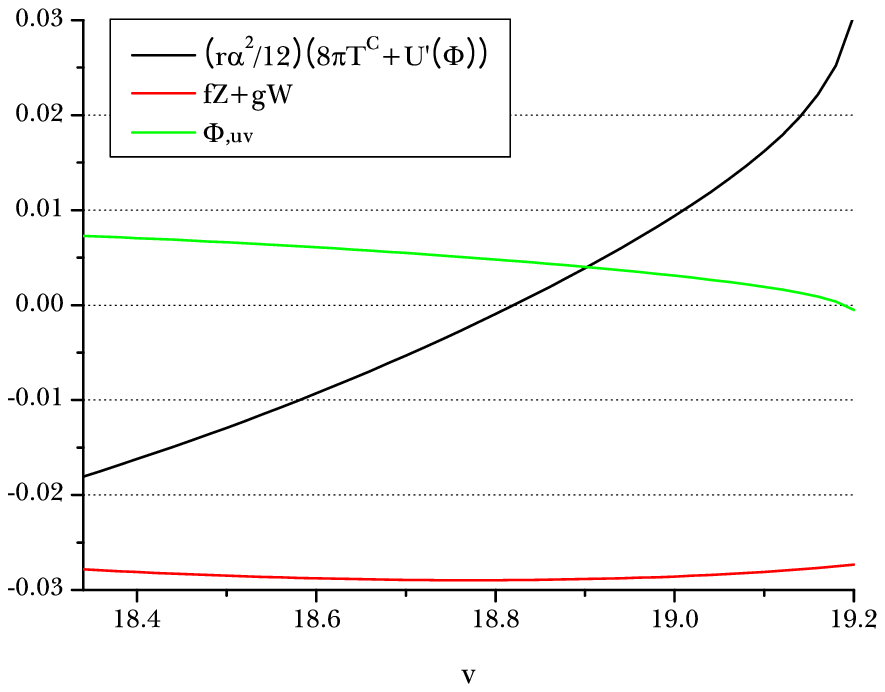}
\includegraphics[scale=1]{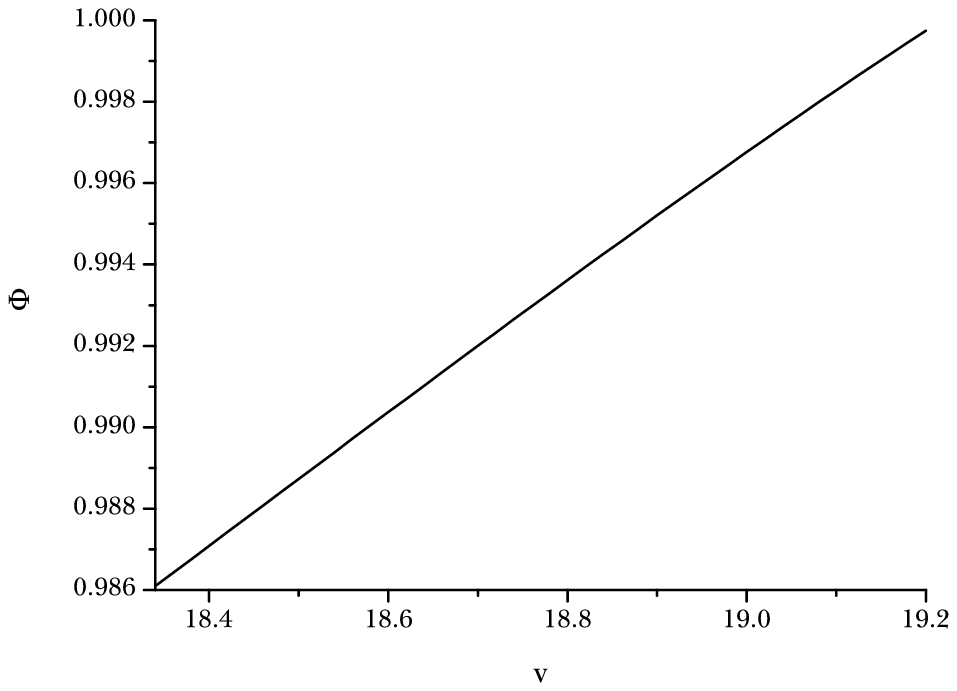}
\caption{\label{fig:fRinducedSb}Some terms $(r\alpha^{2}/12)(8\pi T^{\mathrm{C}}+U')$, $fZ+gW$, and $\Phi_{,uv}$ (upper) and $\Phi$ (lower) as functions of $v$ along $u=15$ for the $A=0.25$ and $\delta=0.5 \pi$ case (Model~$1$). We focus on the behavior near the singularity.}
\end{center}
\end{figure}

Figure~\ref{fig:fRinducedSb} explains what is going on around the singularity. The upper diagram shows three terms for the Brans-Dicke field equation, which is
\begin{eqnarray}
r\Phi_{,uv} = \mathfrak{C} = - \left(fZ + gW\right) - \frac{r\alpha^{2}}{12} \left( 8 \pi T^{\mathrm{C}} + U'(\Phi) \right).
\end{eqnarray}
The first part of the right hand side $(fZ+gW)$ is negative while the second part of the right hand side $(r\alpha^{2}/12)(8\pi T^{\mathrm{C}}+U')$ changes its sign. The latter increases as $v$ increases. This becomes an effective resistance for $\Phi$ that prohibits it to touch $\Phi=1$ (i.e., $\sim U'(\Phi)$). However, $\Phi_{,uv}$ will be solely determined by the sum of the first and the second term, and if the conditions are finely tuned, the first term and the second term can be exactly canceled around $\Phi \simeq 1$. Then
\begin{eqnarray}
\frac{\Phi_{,uv}}{\Phi} \ll 1 \;\;\;\; \mathrm{and} \;\;\;\; \frac{\Phi_{,u}}{\Phi} \sim \frac{\Phi_{,v}}{\Phi} \ll 1
\end{eqnarray}
will hold. (In this calculation, the former is of the order of $10^{-3}$ and the latter is that of $10^{-2}$.) Then $\Phi$ becomes a slowly varying function; as $v$ increases up to order of one, the Brans-Dicke field $\Phi$ varies linearly (lower diagram of Figure~\ref{fig:fRinducedSb}) as a function of $v$. In this case, the increased value $\Delta \Phi$ relative to the scale of $\Phi$ ($\approx 1$) is of the order of $10^{-2}$, and hence all the arguments are consistent.

The physical importance is not easy to interpret. Since the $f(R)$-induced singularity can lie outside a black hole, we can infer that it may have some implications towards astrophysical observations, if we assume $f(R)$ gravity to be the correct theory for our universe. On the other hand, if we believe cosmic censorship, then such $f(R)$ will not be allowed. This crucially depends on the choice of the $f(R)$ function. Moreover, even though a model allows $f(R)$-induced singularity, if we suitably choose the model parameters, then the specific $f(R)$ model will not allow the existence of such naked $f(R)$-induced singularity for astrophysical scales. Therefore, further careful studies are needed.

\subsubsection{Intermediate limit: Oscillating horizons}

Here, we observe $A=0.25$ and $\delta=0.3\pi$ (upper left of Figure~\ref{fig:Model1}). Now, we can see the $f(R)$-induced singularity and the Cauchy horizons, and they are inside the outer apparent horizon. There are some oscillatory apparent horizons inside the outer apparent horizon. This is due to the Brans-Dicke field and the potential. Since the Brans-Dicke field oscillates, the energy-momentum tensor components oscillate, and hence the sign of $r_{,v}$ can oscillate.

\begin{figure}
\begin{center}
\includegraphics[scale=1]{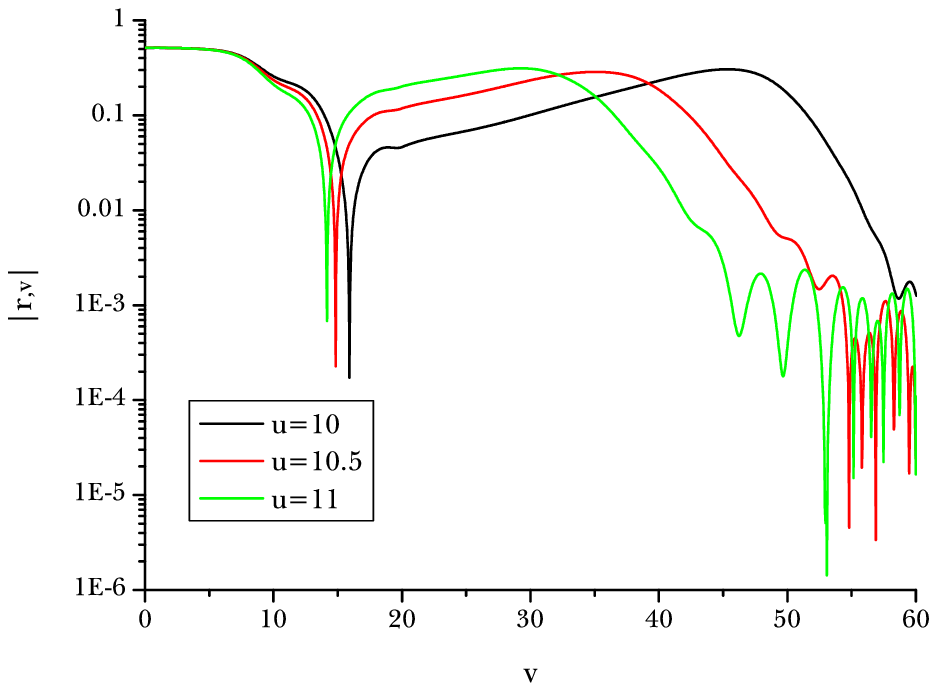}
\caption{\label{fig:G}$|r_{,v}|$ along $u=10, 10.5, 11$ of the $A=0.25$ and $\delta=0.3\pi$ case (Model~$1$). Since $|r_{,v}|$ is plotted on the log scale, one can see that $r_{,v}$ does not tend to zero in $v\rightarrow v_{\textrm{Max}}$ limit.}
\end{center}
\end{figure}

An interesting question is whether such oscillating horizons do converge to form an inner Cauchy horizon in the $v \rightarrow v_{\mathrm{Max}}$ limit\footnote{$v_{\mathrm{Max}}$ is the largest value of the advanced time $v$ inside a charged black hole. For an asymptotically de Sitter space, $v_{\mathrm{Max}}$ should be a finite value. For an asymptotically flat space, $v_{\mathrm{Max}}$ can be infinite.} as in the case of the Einstein gravity or not. Here, $v_{\mathrm{Max}}$ is not infinite, since we consider the black hole in a de Sitter space. Figure~\ref{fig:G} shows that $r_{,v}$ will oscillate, but will never converge to zero. Therefore, in the $v\rightarrow v_{\mathrm{Max}}$ limit, there will be \textit{a Cauchy horizon, but not an inner apparent horizon} inside a charged black hole.

\begin{figure}
\begin{center}
\includegraphics[scale=1]{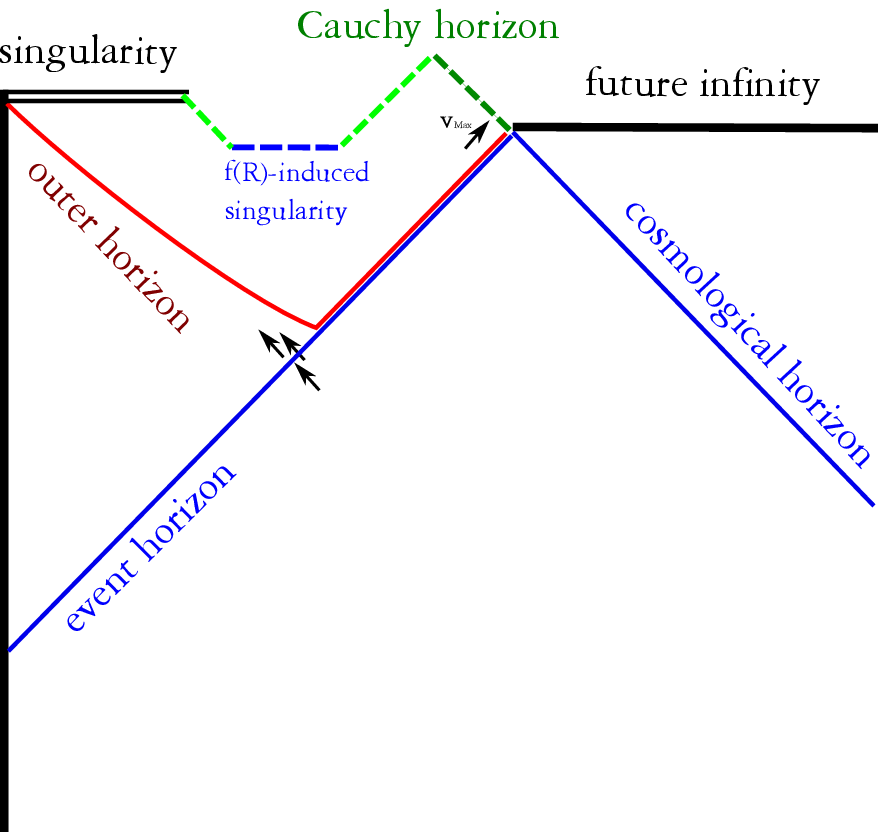}
\caption{\label{fig:diagram}Penrose diagram of a general dynamical charged black hole in $f(R)$ gravity.}
\end{center}
\end{figure}

Finally, we summarize various features of gravitational collapse of the $f(R)$ charged black holes in Figure~\ref{fig:diagram}. We list the following basic and new features of dynamical charged black holes in $f(R)$ gravity:
\begin{itemize}
\item Since $f(R)$ gravity is the $\omega=0$ limit of the Brans-Dicke theory (with a potential), the gravitational collapse will thus push the Brans-Dicke field value to a lower one than the asymptotic value, and this causes the perturbation of the Brans-Dicke field.
\item Perturbations of the Brans-Dicke field can cause the $f(R)$-induced singularity, which in turn will make in-going and out-going Cauchy horizons.
\item There is a space-like singularity and a separation between the space-like singularity and the outer apparent horizon in $v \rightarrow v_{\mathrm{Max}}$ limit, as in the case of the Einstein gravity; however, the $v \rightarrow v_{\mathrm{Max}}$ limit is not an inner apparent horizon rather an in-going null Cauchy horizon.
\end{itemize}

\subsection{Model~$2$: Black holes without $f(R)$-induced singularity}

Due to the $f(R)$-induced singularity, it is difficult to see the inner structure of a black hole. Also, it is important to check whether we can build a model without $f(R)$-induced singularity or not. We find that the $R+(1/2)cR^{2}$ model is a good toy model to observe such a behavior. In this paper, we fix $c=10^{-3}$.

\begin{figure}
\begin{center}
\includegraphics[scale=0.3]{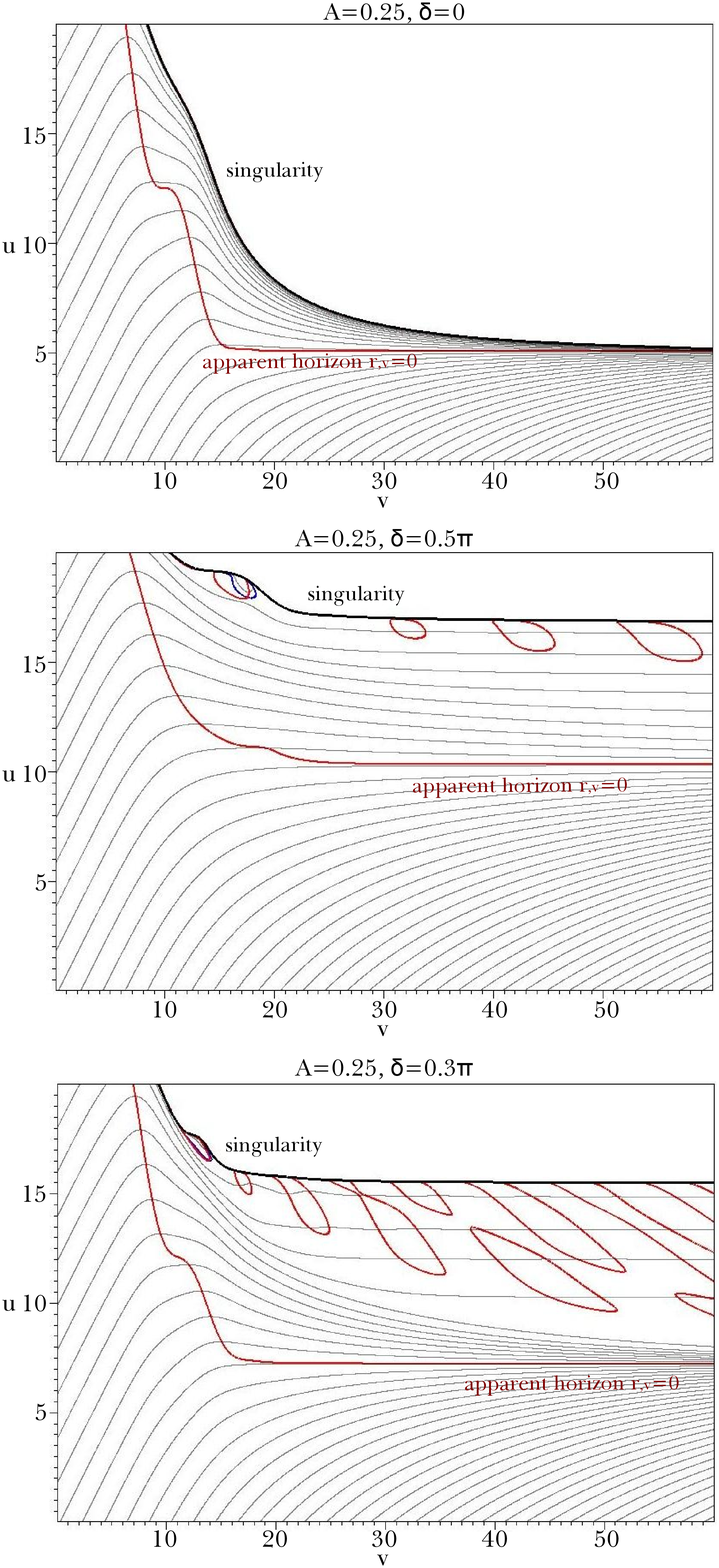}
\caption{\label{fig:Model2}Results of Model~$2$: the neutral limit (top), the extreme limit (middle), and the intermediate limit (bottom).}
\end{center}
\end{figure}

\subsubsection{Neutral limit}

First, we want to reproduce a neutral black hole from the Model~$2$ (top of Figure~\ref{fig:Model2}). In this case, we can again see the space-like singularity and the outer apparent horizon.

\subsubsection{Near extreme limit}

If we choose $\delta=0.5 \pi$, we can get a near extreme black hole for the Model~$2$ (middle of Figure~\ref{fig:Model2}). Here, we see a drastic difference as compared to Model~$1$. We can observe a space-like singularity and an outer apparent horizon. In addition, there is a clear separation between the singularity and the outer horizon inside the black hole. This is quite similar to that of a near extreme charged black hole in the Einstein gravity \cite{Yeom1}\cite{Yeom2}\cite{Yeom3}. One interesting difference is there are oscillatory $r_{,v}$ horizons near the singularity, due to the term coming from the Brans-Dicke field. In pure Einstein gravity with charged scalar fields, since there is no source to violate the null energy condition, we do not get any such behavior.

\subsubsection{Intermediate limit}

Finally, we observe the intermediate charged black hole case (bottom of Figure~\ref{fig:Model2}). Again, there is a space-like singularity and an outer apparent horizon. In this case, the oscillatory behavior of $r_{,v}$ horizons is much clear. One can easily compare the size of the outer apparent horizon for each result. Due to the charge repulsion, the size of black holes become larger and larger, as one decreases the charge \cite{Yeom3}.

Therefore, as conclusion for the study of $R+(1/2)cR^{2}$ model, we can notice that the existence of the $f(R)$-induced singularity is not a necessary property of the $f(R)$ gravity.

\subsubsection{Question: When does the $f(R)$-induced singularity appear?}

At this stage, the natural question that one might address is the following: when does the $f(R)$-induced singularity appear? In other words, what are the conditions that induce such a singularity? To give a rigorous answer for the question, we need to check various kinds of $f(R)$ models and hence it goes beyond the scope of the present paper. However, we can comment on some of the basic intuitions.

Let us assume $f(R) = R + \mathrm{const.} + \mathcal{O}(R^{-1}) + ...$ for the $R \rightarrow \infty$ limit. This implies that the model approaches the Einstein theory in the $R \rightarrow \infty$ limit and hence the Brans-Dicke field should be a constant in the limit: $\Phi \rightarrow \Phi_{0}$. Note that, there is a relation between $R$ and $\Phi$ by $V'(\Phi) = R$. Therefore, in the $R \rightarrow \infty$ limit, $V'(\Phi_{0}) = R \rightarrow \infty$; at the same time, $U'(\Phi_{0}) = 2f(R) - R f'(R) \simeq R \rightarrow \infty$. Also, we can check that $U(\Phi_{0}) = \int^{\infty} (2f(R)-Rf'(R)) f''(R) dR \propto \mathcal{O}(R^{-1})$. Therefore, $U(\Phi_{0})$ is finite and the effective potential $U(\Phi)$ has a \textit{cusp} at $\Phi_{0}$. Then, the observation of the $f(R)$-induced singularity is possible.

The existence of $f(R)$-induced singularity is related to the existence of a \textit{cusp} in the effective potential $U(\Phi)$ due to a certain $f(R)$ model. We can conclude that, \textit{the existence of $f(R)$-induced singularity is possible if $f(R) \sim R$ for the large $R$ limit}, although it is not a general behavior of the $f(R)$ gravity. 

\subsection{Mass inflation in $f(R)$ gravity}

We can define the null geodesics (here, our convention is $[u,v,\theta,\varphi]$) as
\begin{eqnarray}
l^{\mu}&=&\frac{\sqrt{2}}{\alpha}(0,1,0,0),\\
n^{\mu}&=&\frac{\sqrt{2}}{\alpha}(1,0,0,0),
\end{eqnarray}
so that $l^{\mu}l_{\mu}=n^{\mu}n_{\mu}=0$ and $l^{\mu}n_{\mu}=-1$. Then, any time-like geodesic $t^{\mu}$ can be decomposed as
\begin{eqnarray}
t^{\mu}= a l^{\mu} + b n^{\mu}
\end{eqnarray}
with the constraint $ab=1/2$. Hence, any time-like observer will see the local energy density to be
\begin{eqnarray}
T_{\mu\nu} t^{\mu}t^{\nu}=\frac{2}{\alpha^{2}} \left( b^{2} T_{uu} + a^{2} T_{vv} \right) + \frac{2}{\alpha^{2}}T_{uv}.
\end{eqnarray}
Therefore, if for some case $T_{uu}/\alpha^{2}$, $T_{vv}/\alpha^{2}$, or $T_{uv}/\alpha^{2}$ increases exponentially, for general choices of $a$ and $b$, we can sure that it is the signature for the evidence of mass inflation.

\begin{figure}
\begin{center}
\includegraphics[scale=0.8]{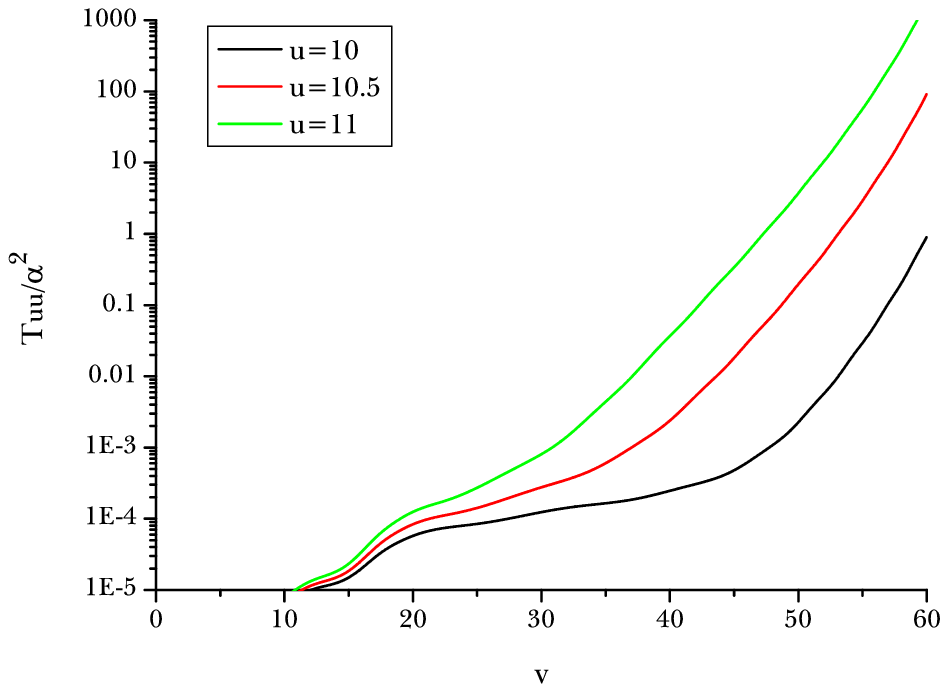}
\includegraphics[scale=0.8]{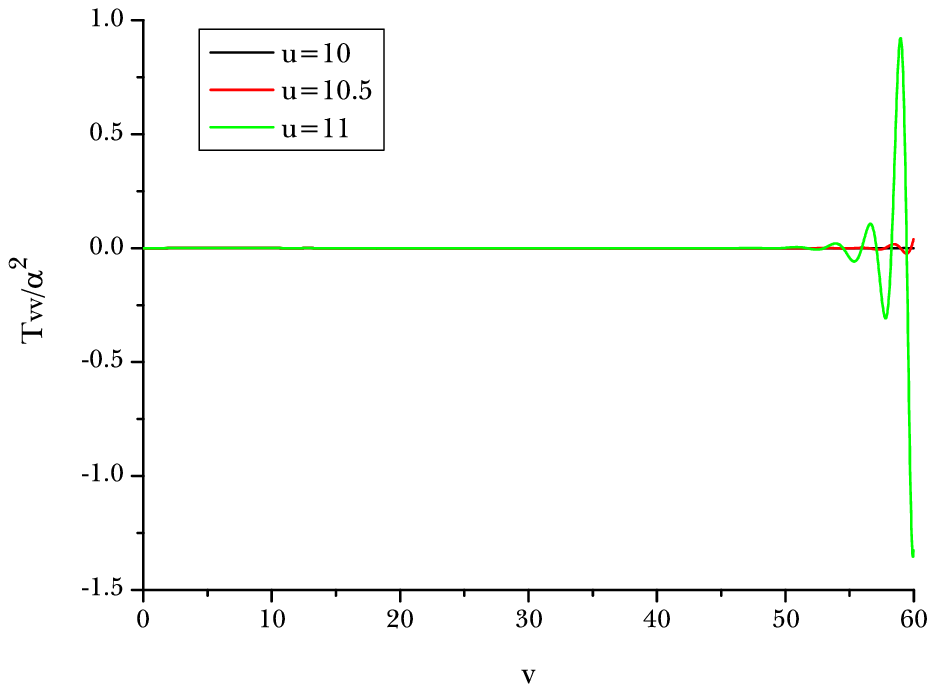}
\includegraphics[scale=0.8]{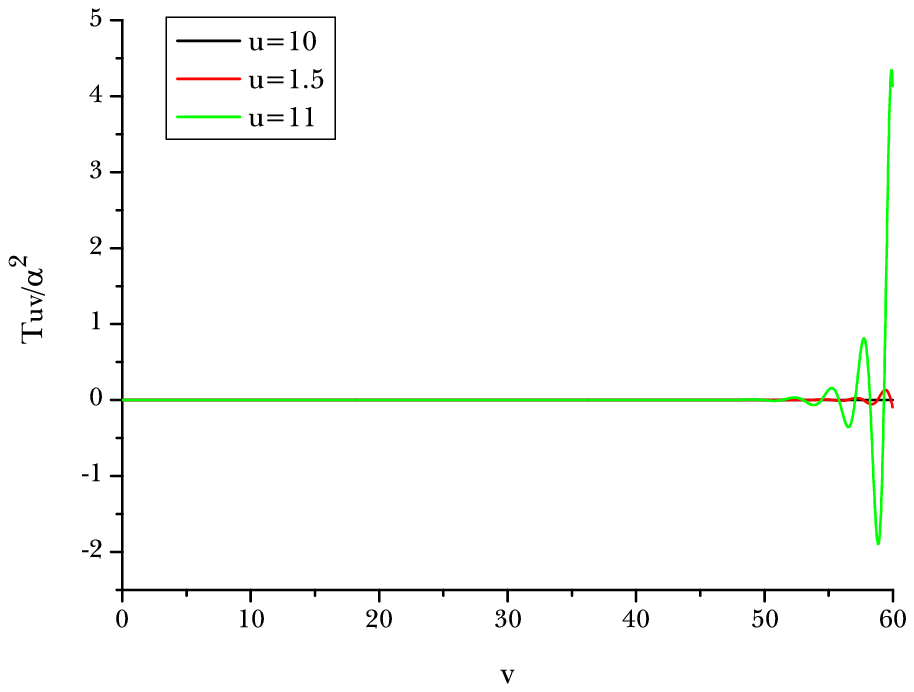}
\caption{\label{fig:Tuu_1}$T_{uu}/\alpha^{2}$ (upper), $T_{vv}/\alpha^{2}$ (middle), and $T_{uv}/\alpha^{2}$ (lower) of the $A=0.25$ and $\delta=0.3\pi$ case (Model~$1$). $T_{uu}/\alpha^{2}$ is plotted on the log scale; this shows an exponential increase in a component of the energy-momentum tensor as $\sim \exp \kappa v$, where $\kappa$ is a constant.}
\end{center}
\end{figure}

Figure~\ref{fig:Tuu_1} shows $T_{uu}/\alpha^{2}$ (upper), $T_{vv}/\alpha^{2}$ (middle), and $T_{uv}/\alpha^{2}$ (lower) for the $A=0.25$ and $\delta=0.3\pi$ case of the Model~$1$. The former increases exponentially, while the latter does not show any such behavior. Therefore, it is quite reasonable to conclude from this study that $f(R)$ gravity has mass inflation.

\begin{figure}
\begin{center}
\includegraphics[scale=1]{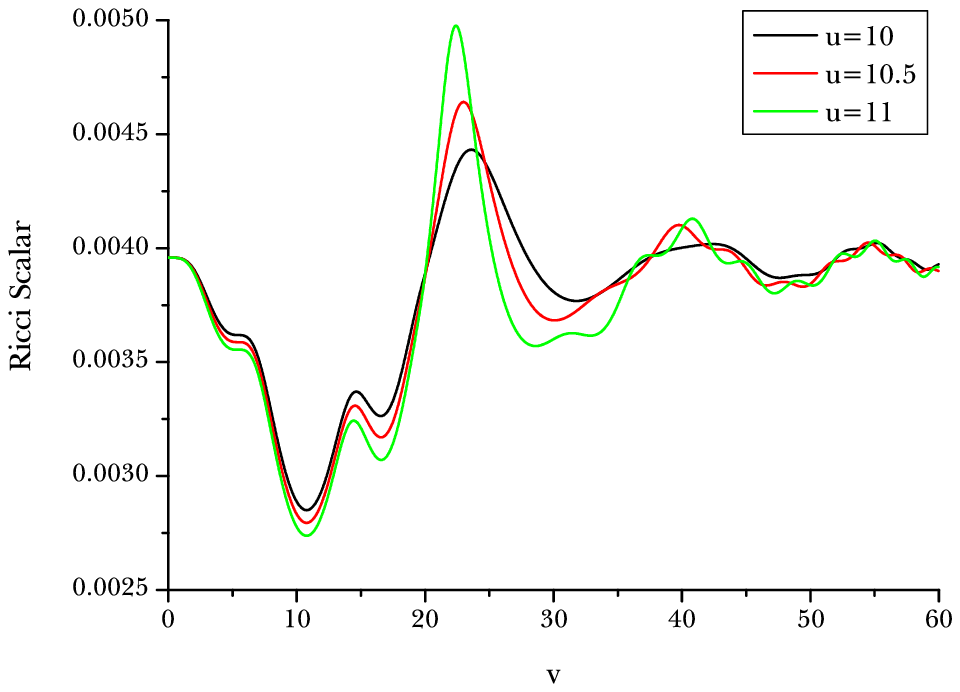}
\caption{\label{fig:Ricci_1}Ricci scalar $R$ for the $A=0.25$ and $\delta=0.3\pi$ case (Model~$1$).}
\end{center}
\end{figure}

However, for the same case, we can check the behavior of the Ricci scalar $R$ (Figure~\ref{fig:Ricci_1}) and we can easily notice that it is finite. This is due to the potential of the Brans-Dicke field. If the Brans-Dicke field can be confined around the local minimum, then the Ricci scalar $R$ should be of the order of $2V/\Phi \sim 0.004$. This is not entirely trivial; as we saw in the $f(R)$-induced singularity case, the Ricci scalar can diverge via complex interactions between the matter and gravity. However, we see that \textit{mass inflation does not induce the Ricci scalar to diverge}. If we compare the result with the Einstein case (Figure~16 in \cite{Yeom2}), we see that it is quite a distinct property from that of the Einstein gravity.

\begin{figure}
\begin{center}
\includegraphics[scale=0.8]{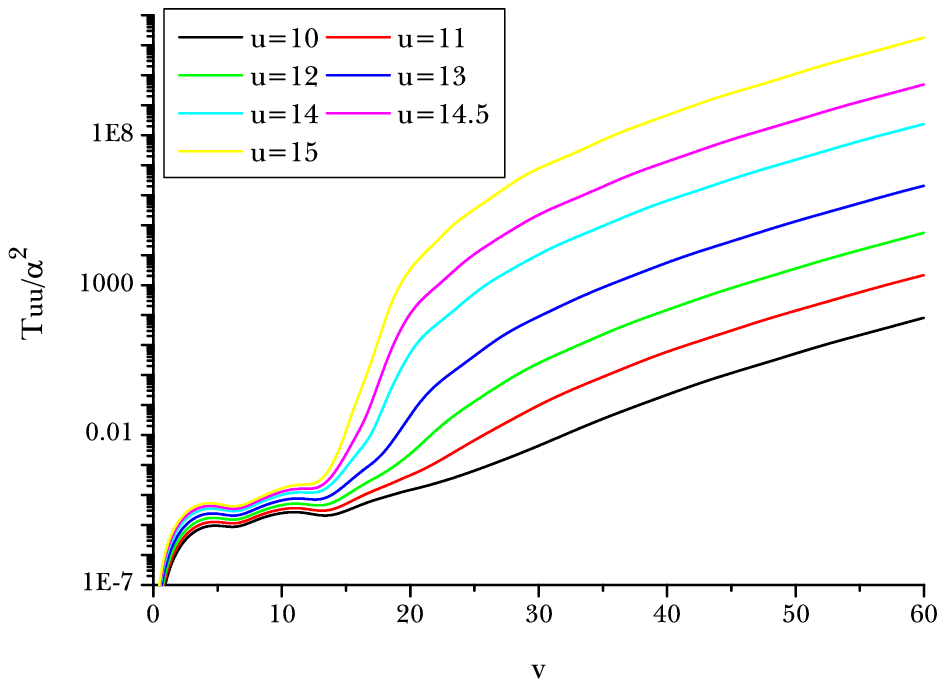}
\includegraphics[scale=0.8]{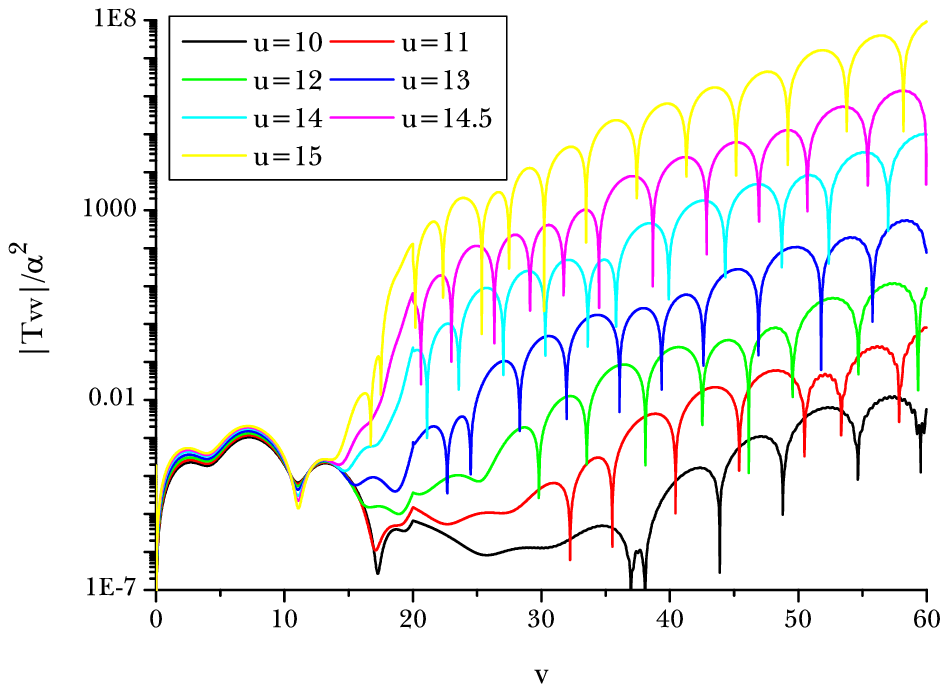}
\includegraphics[scale=0.8]{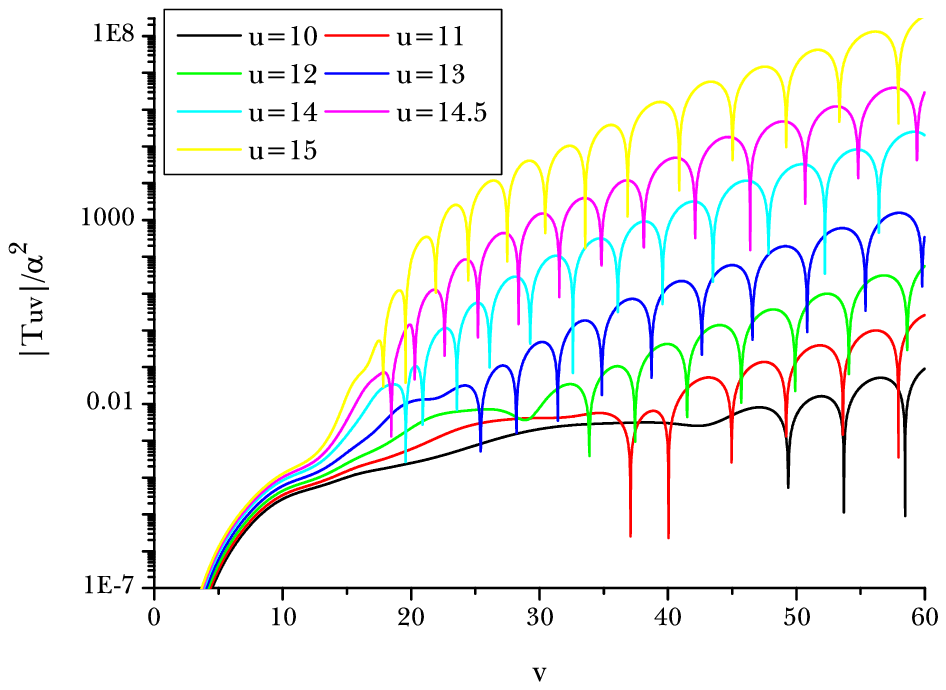}
\caption{\label{fig:Tuu_2}$T_{uu}/\alpha^{2}$ (upper), $|T_{vv}|/\alpha^{2}$ (middle), and $|T_{uv}|/\alpha^{2}$ (lower) of the $A=0.25$ and $\delta=0.3\pi$ case (Model~$2$). All figures are plotted on the log scale; this shows an exponential increase in a component of the energy-momentum tensor as $\sim \exp \kappa v$, where $\kappa$ is a constant.}
\end{center}
\end{figure}

The Model~$2$ gives similar results. Figure~\ref{fig:Tuu_2} shows $T_{uu}/\alpha^{2}$ (upper), $|T_{vv}|/\alpha^{2}$ (middle), and $|T_{uv}|/\alpha^{2}$ (lower) on a log scale. The tendency to linearly increase shows the existence of mass inflation clearly. 

\begin{figure}
\begin{center}
\includegraphics[scale=1.1]{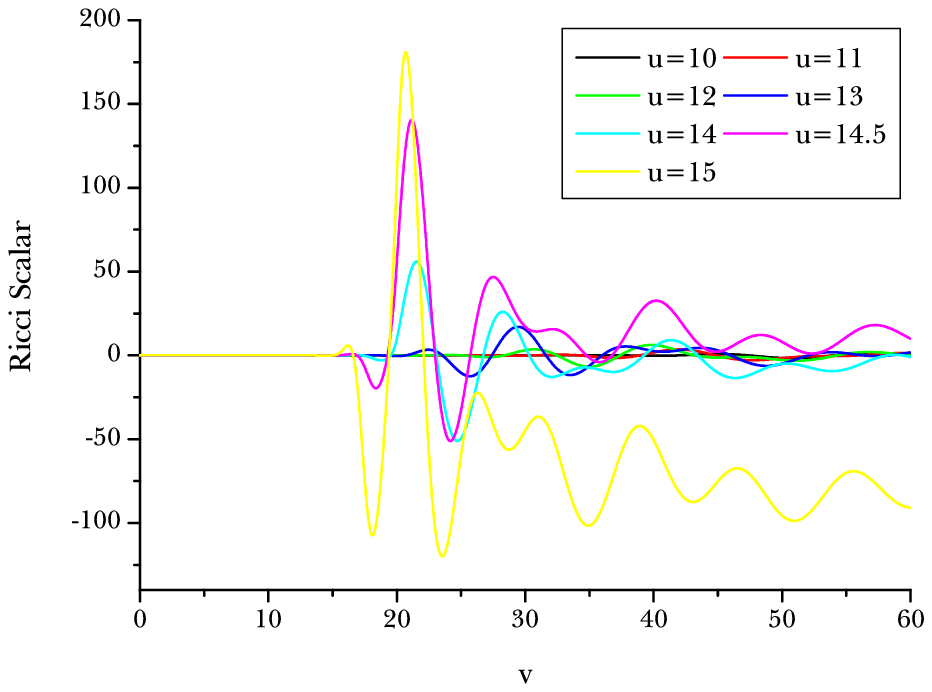}
\caption{\label{fig:Ricci_2}Ricci scalar $R$ for the $A=0.25$ and $\delta=0.3\pi$ case (Model~$2$).}
\end{center}
\end{figure}

\begin{figure}
\begin{center}
\includegraphics[scale=1.1]{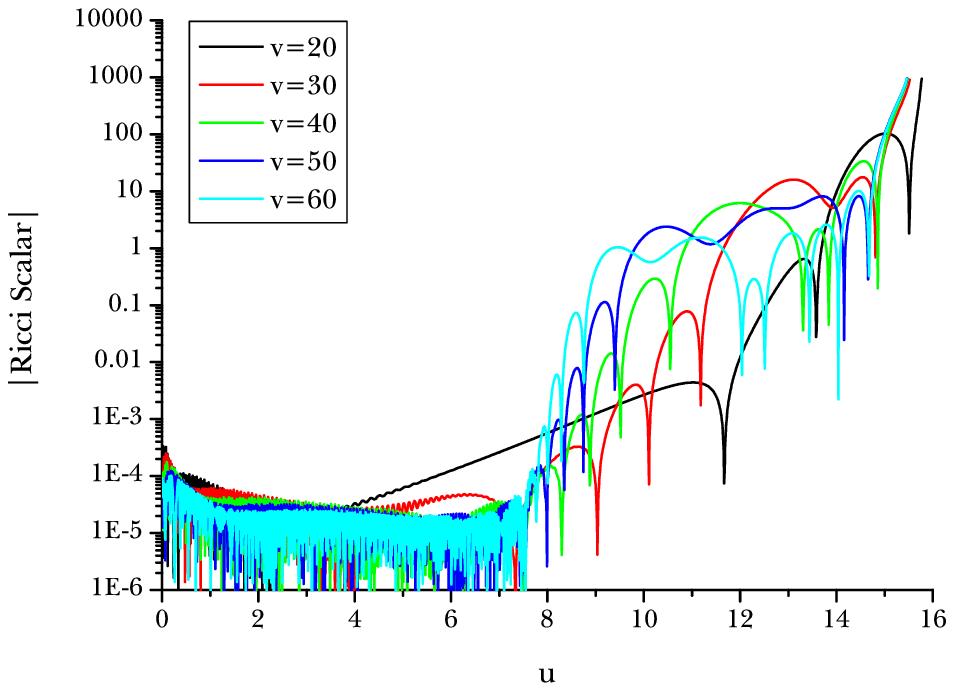}
\caption{\label{fig:Ricci_2_u}Ricci scalar $|R|$ for the $A=0.25$ and $\delta=0.3\pi$ case (Model~$2$). The $|R|$ axis is the log scale. In this case, as $u$ increases, there is a tendency of increase in the Ricci scalar.}
\end{center}
\end{figure}

However, the Ricci scalar in Figure~\ref{fig:Ricci_2} is not affected by mass inflation. It oscillates around an equilibrium point. One important note is that, although the Ricci scalar oscillates around an equilibrium position along a fixed $u$, there is a tendency that it increases as $u$ increases. Figure~\ref{fig:Ricci_2_u} shows some detailed results for $|R|$ along some fixed $v$. As $u$ increases, the amplitude of oscillations of $R$ tends to increase, but it remains sufficiently small unless it approaches the central singularity. The scale of Ricci scalar is determined by $c$, as
\begin{eqnarray}
R = V'(\Phi)=\frac{1}{c}(\Phi-1).
\end{eqnarray}
This implies that as the higher curvature correction term becomes more and more dominant (as $c$ increases), the Ricci scalar $R$ will slowly increase.

To summarize the findings for the Model~$1$ and Model~$2$, we can conclude the following three things:
\begin{itemize}
\item There \textit{is} mass inflation, such that an in-falling observer will measure exponential divergence of some of energy-momentum tensor components. Then, $v\rightarrow v_{\mathrm{Max}}$ limit may be a null curvature singularity, similar to the Einstein case.
\item The Ricci scalar is bounded by a certain value along the out-going null direction.
\item The Ricci scalar increases along the in-going null direction; but, the in-going observer will quickly reach the space-like central singularity.
\end{itemize}


\section{\label{sec:dis}Discussion}

In this paper, we studied the gravitational collapse and dynamical behavior of a charged $f(R)$ black hole. To satisfy the cosmological stability criterion, we chose some special models of $f(R)$: one of them is the Starobinsky model, the other being the simplest $R+(1/2)cR^{2}$ model.

The higher curvature corrections will change the basic features of the dynamical charged black holes from that of the original Einstein gravity. Since $f(R)$ gravity is the $\omega=0$ limit of the Brans-Dicke theory (with a potential), the gravitational collapse will push the Brans-Dicke field to a lower value than the asymptotic one, and this causes the perturbation of the Brans-Dicke field. This was already known from the authors' previous work \cite{Hwang:2010aj}; however, the perturbations of the Brans-Dicke field can cause a $f(R)$-induced singularity, and \textit{this is indeed a new observation for the $f(R)$ gravity}. This $f(R)$-induced singularity will in turn lead to in-going and out-going Cauchy horizons. There is a space-like singularity and a separation between the space-like singularity and the outer apparent horizon in the $v \rightarrow v_{\mathrm{Max}}$ limit, as in the case of Einstein gravity; however, \textit{the $v \rightarrow v_{\mathrm{Max}}$ limit may not be an inner apparent horizon, and will be just an in-going null Cauchy horizon.}

The main motivation of this paper is to study mass inflation in $f(R)$ gravity. We observed some interesting features. First, there \textit{is} mass inflation and an in-falling observer will measure exponential divergence in some energy-momentum tensor components similar to the Einstein case. However, interestingly, the Ricci scalar is bounded by a certain value along the out-going null direction. 

An out-going observer will never see a geodesically incomplete central singularity; rather, the observer will just see a curvature singularity caused by mass inflation with non-zero area. What is indeed a new feature is that \textit{the Ricci scalar is bounded for the observer}. This is due to the higher curvature correction terms of the $f(R)$ gravity, and also is due to the stable local minimum of the effective potential $U(\Phi)$. Thus, we can conclude that cosmologically `stable' `higher curvature corrections' can indeed hold the Ricci scalar to be of a finite value, even in the presence of mass inflation.

Let us emphasize our results again. The action of the present paper is
\begin{eqnarray}\label{eq:action}
S = \int dx^{4} \sqrt{-g} \left[ \frac{f(R)}{16 \pi} + \mathcal{L}_{\mathrm{matter}} + f_{2}(R_{\mu\nu\rho\sigma}^{2}) + f_{3}(R_{\mu\nu}^{2}) + ... \right].
\end{eqnarray}
We only use the first and the second leading terms, since we assume that contributions of $f_{2}, f_{3}, ...$ are sufficiently smaller than the leading terms.
Then,
\begin{enumerate}
\item The $f(R)$ sector is finite even in the presence of mass inflation.
\item The $\mathcal{L}_{\mathrm{matter}}$ sector will increase larger than the $f(R)$ sector. This is easy to check: if there is a kinetic term of an arbitrary scalar field $\psi$, then it will be proportional to $1/\alpha^{2}$, where $\alpha$ exponentially decreases in the presence of mass inflation. This shows the instability of the matter sector of the Lagrangian.
\item Some of the energy-momentum tensor components are related to some of the curvature components by the Einstein equation. Therefore, some terms of higher order curvature corrections $f_{2}, f_{3}, ...$ can be larger and larger in the presence of mass inflation. This may violate the original assumption of the action in Equation~(\ref{eq:action}).
\item We can conjecture that if we equally consider all the higher order curvature corrections $f_{2}, f_{3}, ...$ and if we impose the cosmological stability\footnote{Here, the meaning of the cosmological stability should be well-defined. If the general action Equation~(\ref{eq:action}) is equivalent to the Einstein gravity with a number of scalar, vector, or tensor fields, then we can impose the stability to the scalar, vector, or tensor fields; they should be asymptotically stable up to field perturbations. However, in the present paper, we cannot justify whether we can impose the concept for the most general case. But, at least, we can say that this is a reasonable generalization. For this point, we thank to the anonymous referee.} for the entire action, then all the curvature components can be finite even in the presence of mass inflation.
\item Then, the observed energy density of an in-falling observer near the Cauchy horizon can be finite even in a charged black hole. This can be a resolution of the mass inflation curvature singularity.
\end{enumerate}
Note that $1$, $2$, and $3$ are confirmed by the present paper, and $4$ and $5$ are our conjecture.

In conclusion, we have a promising, yet brave, conjecture, which is as follows: \textit{if we extend the action to include not only a function of the Ricci scalar, but also functions of more complex combinations of curvature, and if we require cosmological stability for the entire action, then all the curvature components and the observed local energy densities should be finite around the Cauchy horizons of charged black holes}. Then, we can trust the original action, as all the components of the action may be bounded by a certain value, and we can obtain a self-consistent gravitational dynamics in the presence of inner Cauchy horizons. This will shed some light on the study of regular black holes. Classical regular black holes have inner horizons in general, and they cannot be made free from the mass inflation instability. However, if higher order curvature corrections can make all the curvature components to be bounded, then it will be possible to understand the consistency condition for the action of a certain model, and it will for sure help us to understand more realistic regular black hole models.

\section*{Acknowledgment}
DH would like to thank Ewan Stewart. DY would like to thank Raju Roychowdhury and Chaitali Roychowdhury for the careful English revision. DY and BHL are supported by the National Research Foundation of Korea(NRF) grant funded by the Korea government(MEST) through the Center for Quantum Spacetime(CQUeST) of Sogang University with grant number 2005-0049409. DH is supported by Korea Research Foundation grants (KRF-313-2007-C00164, KRF-341-2007-C00010) funded by the Korean government (MOEHRD) and BK21.

\newpage

\section*{\label{sec:appb}Appendix. Consistency and convergence tests}

In this appendix, we report on the convergence and consistency tests for our simulations. We used the $A=0.25$ and $\delta=0.3\pi$ case of the Starobinsky model, since this case includes all the interesting features of the present paper.

\begin{figure}
\begin{center}
\includegraphics[scale=1]{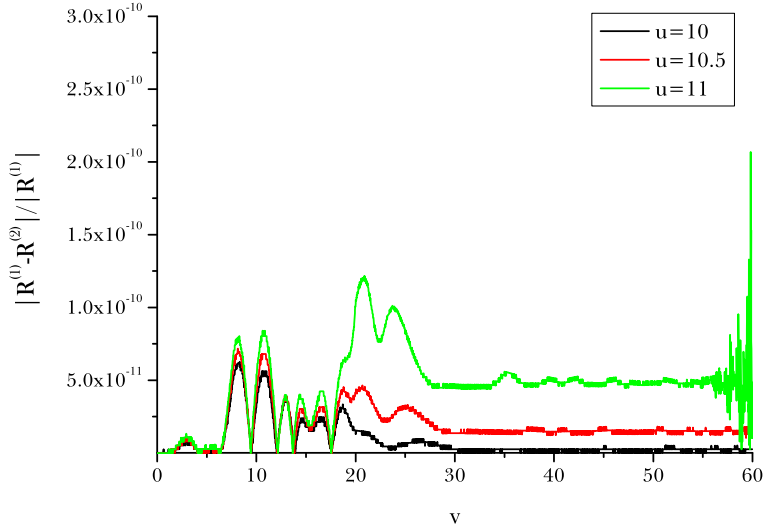}
\caption{\label{fig:consistency1}Consistency test for $|R^{(1)}-R^{(2)}|/|R^{(1)}|$ around $u=10, 10.5, 11$, for the $A=0.25$ and $\delta=0.3\pi$ of the Starobinsky model.}
\end{center}
\end{figure}

For consistency, we can check various relations, but the most important non-trivial test for the $f(R)$ gravity will be the Ricci scalar. The definition of the Ricci scalar is found in Equation~(\ref{eq:Ricci}), while we calculated it by using Equations~(\ref{eq:R1}) and (\ref{eq:R2}). We call the former $R^{(1)}$ and the latter $R^{(2)}$, and checked $|R^{(1)}-R^{(2)}|/|R^{(1)}|$ around $u=10, 10.5, 11$, where mass inflation occurs. Figure~\ref{fig:consistency1} shows that the differences are less than $10^{-8}$~\% and hence it is sufficiently small.

\begin{figure}
\begin{center}
\includegraphics[scale=1]{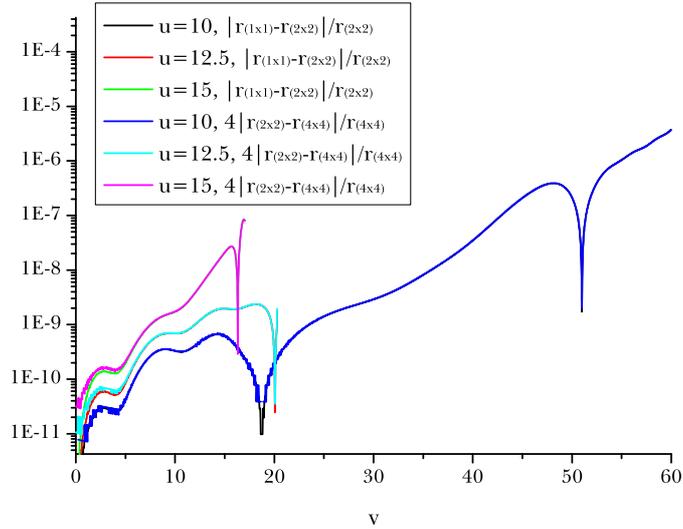}
\caption{\label{fig:convergence}Convergence test: $|r_{(1\times1)}-r_{(2\times2)}|/r_{(2\times2)}$ and $4|r_{(2\times2)}-r_{(4\times4)}|/r_{(4\times4)}$ do coincide around $u=10, 12.5, 15$, for the $A=0.25$ and $\delta=0.3\pi$ of the Starobinsky model. This shows the second order convergence.}
\end{center}
\end{figure}

For convergence, we compared finer simulations: $1\times1$, $2\times2$, and $4\times4$ times finer for some important slices (during mass inflation and near the $f(R)$ induced singularity).
In Figure~\ref{fig:convergence}, we see that the difference between the $1\times1$ and $2\times2$ times finer cases is $4$ times the difference between the $2\times2$ and $4\times4$ times finer cases,
and thus our simulation converges to second order. The numerical error is $\lesssim 10^{-3}\%$.



\newpage

\begin{thebibliography}{200}

\bibitem{Gasperini:2007zz}
  M.~Gasperini, {\it ``Elements of string cosmology,''} Cambridge, Cambridge University Press (2007).

\bibitem{Brans:1961sx}
  C.~Brans and R.~H.~Dicke,
  Phys.\ Rev.\  {\bf 124}, 925 (1961).

\bibitem{Magnano:1993bd}
  G.~Magnano and L.~M.~Sokolowski,
  Phys.\ Rev.\  D {\bf 50}, 5039 (1994)
  [arXiv:gr-qc/9312008].

\bibitem{Sotiriou:2008rp}
  T.~P.~Sotiriou and V.~Faraoni,
  Rev.\ Mod.\ Phys.\  {\bf 82}, 451 (2010)
  [arXiv:0805.1726 [gr-qc]];\\
  S.~Nojiri and S.~D.~Odintsov,
  eConf {\bf C0602061}, 06 (2006)
  [Int.\ J.\ Geom.\ Meth.\ Mod.\ Phys.\  {\bf 4}, 115 (2007)]
  [arXiv:hep-th/0601213].


\bibitem{Hawking:1973uf}
  S.~W.~Hawking and G.~F.~R.~Ellis, {\it ``The large scale structure of space-time,''} Cambridge, Cambridge University Press (1973).

\bibitem{Frolov:1988vj}
  V.~P.~Frolov, M.~A.~Markov and V.~F.~Mukhanov,
  Phys.\ Rev.\  D {\bf 41}, 383 (1990).

\bibitem{Hossenfelder:2009fc}
  S.~Hossenfelder, L.~Modesto and I.~Premont-Schwarz,
  Phys.\ Rev.\  D {\bf 81}, 044036 (2010)
  [arXiv:0912.1823 [gr-qc]].

\bibitem{Poisson:1990eh}
  E.~Poisson and W.~Israel,
  Phys.\ Rev.\  D {\bf 41}, 1796 (1990).

\bibitem{Brown:2011tv}
  E.~G.~Brown, R.~B.~Mann and L.~Modesto,
  arXiv:1104.3126 [gr-qc].

\bibitem{Yeom:2009zp}
  D.~Yeom and H.~Zoe,
  Int.\ J.\ Mod.\ Phys.\  A {\bf 26}, 3287 (2011)
  [arXiv:0907.0677 [hep-th]].

\bibitem{Ashtekar:2005cj}
  A.~Ashtekar and M.~Bojowald,
  Class.\ Quant.\ Grav.\  {\bf 22}, 3349 (2005)
  [arXiv:gr-qc/0504029].

\bibitem{Yeom1}
  J.~Hansen, D.~Hwang and D.~Yeom, JHEP {\bf 0911}, 016 (2009) [arXiv:0908.0283 [gr-qc]]; \\
  D.~Hwang and D.~Yeom,
  Class.\ Quant.\ Grav.\  {\bf 28}, 155003 (2011)
  [arXiv:1010.3834 [gr-qc]]; \\
  D.~Hwang, H.~Kim and D.~Yeom,
  arXiv:1105.1371 [gr-qc].

\bibitem{Yeom2}
  S.~E.~Hong, D.~Hwang, E.~D.~Stewart and D.~Yeom,
  Class.\ Quant.\ Grav.\  {\bf 27}, 045014 (2010) [arXiv:0808.1709 [gr-qc]].

\bibitem{Yeom3}
  D.~Hwang and D.~Yeom,
  Phys.\ Rev.\  D {\bf 84}, 064020 (2011)
  [arXiv:1010.2585 [gr-qc]].


\bibitem{Hwang:2010aj}
  D.~Hwang and D.~Yeom,
  Class.\ Quant.\ Grav.\  {\bf 27}, 205002 (2010)
  [arXiv:1002.4246 [gr-qc]].

\bibitem{Avelino:2009vv}
  P.~P.~Avelino, A.~J.~S.~Hamilton and C.~A.~R.~Herdeiro,
  Phys.\ Rev.\  D {\bf 79}, 124045 (2009) [arXiv:0904.2669 [gr-qc]].

\bibitem{nr}
  W.~H.~Press, S.~A.~Teukolsky, W.~T.~Vetterling and B.~P.~Flannery, {\it ``Numerical Recipes: The Art of Scientific Computing,'' 3rd ed.}
  Cambridge, Cambridge University Press (2007).

\bibitem{Starobinsky:2007hu}
  A.~A.~Starobinsky,
  JETP Lett.\  {\bf 86}, 157 (2007)
  [arXiv:0706.2041 [astro-ph]].



\end{thebibliography}
\end{document}